\documentclass[12pt]{article}

\usepackage{amsfonts, bm, amsmath, color, natbib, rotating, amsthm, subfigure, lscape, multicol, epstopdf, verbatim, hyperref} 

\usepackage{JASA_manu}
\bibpunct{(}{)}{;}{a}{}{,}

\input psfig.tex

\def\log{\hbox{log}}

\def\boxit#1{\vbox{\hrule\hbox{\vrule\kern6pt
          \vbox{\kern6pt#1\kern6pt}\kern6pt\vrule}\hrule}}

\def\bse{\begin{eqnarray*}}
\def\ese{\end{eqnarray*}}
\def\be{\begin{eqnarray}}
\def\ee{\end{eqnarray}}
\def\bq{\begin{equation}}
\def\eq{\end{equation}}
\def\bse{\begin{eqnarray*}}
\def\ese{\end{eqnarray*}}

\long\def\symbolfootnote[#1]#2{\begingroup%
\def\thefootnote{\fnsymbol{footnote}}\footnote[#1]{#2}\endgroup}

\newcommand{\bS}{{\utwi{S}}}
\newcommand{\s}{{{S}}}
\newcommand{\e}{{\utwi{e}}}
\newcommand{\w}{{\utwi{W}}}
\newcommand{\z}{{{Z}}}
\newcommand{\y}{{{Y}}}
\newcommand{\Y}{{\utwi{Y}}}
\newcommand{\X}{{\utwi{X}}}
\newcommand{\x}{{\utwi{x}}}
\newcommand{\q}{{\utwi{Q}}}

\newcommand{\bO}{{\utwi{O}}}
\newcommand{\A}{{\utwi{A}}}

\newcommand{\m}{{\utwi{M}}}
\newcommand{\I}{{\utwi{I}}}
\newcommand{\p}{{\utwi{P}}}

\newcommand{\V}{{\cal I}}
\newcommand{\T}{{\theta}}
\newcommand{\R}{{\cal R}}
\newcommand{\ba}{{\utwi{a}}}
\newcommand{\bb}{{\utwi{b}}}
\newcommand{\Ti}{{\psi}}
\newcommand{\bt}{{{t}}}

\newcommand{\utwi}[1]{\mbox{\boldmath $ #1$}}

\newcommand{\argmin}{\operatornamewithlimits{argmin}}

\newtheorem{thm}{\normalfont{Theorem}}[section]
\newtheorem{asm}[thm]{\normalfont{Assumption}}
\newtheorem{lem}[thm]{\normalfont{Lemma}}

\begin{document}
\thispagestyle{empty}
\baselineskip=28pt 

\begin{center}
{\LARGE{\bf Independent Component Analysis \\ via Distance Covariance}} 
\end{center}

\baselineskip=20pt 
\vskip 5mm
\begin{center}
David S. Matteson and Ruey S. Tsay
\symbolfootnote[1]{\baselineskip=10pt 
Matteson is Assistant Professor, 
Department of Statistical Science,
Cornell University,
1196 Comstock Hall,
Ithaca, NY 14853
(Email: \href{mailto:matteson@cornell.edu}{matteson@cornell.edu}; Web: \url{http://www.stat.cornell.edu/\~matteson/}).
Tsay is H.G.B. Alexander Professor of Econometrics \& Statistics,
Booth School of Business, University of Chicago,
5807 South Woodlawn Avenue,
Chicago, IL 60637
(Email: \href{mailto:ruey.tsay@chicagobooth.edu}{ruey.tsay@chicagobooth.edu}; Web: \url{http://faculty.chicagobooth.edu/ruey.tsay/}).} 
\vskip 5mm
\today  
\end{center}


\begin{center}
{{ABSTRACT}}
\end{center}
\baselineskip=18pt

This paper introduces a novel statistical framework for independent component analysis (ICA) of multivariate data. 
We propose methodology for estimating and testing the existence of mutually independent components for a given dataset, 
and a versatile resampling-based procedure for inference.
Independent components are estimated by combining a nonparametric probability integral transformation with a generalized nonparametric whitening method that simultaneously minimizes all forms of dependence among the components. 
$U$-statistics of certain Euclidean distances between sample elements are combined in succession to construct a 
statistic for testing the existence of
mutually independent components. 
The proposed measures and tests are based on both necessary and sufficient conditions for mutual independence.
When independent components exist, one may apply univariate analysis to study or model each component separately. 
Univariate models may then be combined to obtain a multivariate model for the original observations. 
We prove the consistency of our estimator under minimal regularity conditions 
without assuming the existence of independent components \emph{a priori}, and 
all assumptions are placed on the observations directly, not on the latent components.
We demonstrate the improvements of the proposed method over competing methods in simulation studies. 
We apply the proposed ICA approach to two real examples and contrast it with principal component analysis.


\baselineskip=18pt

\baselineskip=18pt
\par\vfill\noindent
\par\medskip\noindent
{\bf Short title: Independent Component Analysis}

\clearpage\pagebreak\newpage
\pagenumbering{arabic}
\section{Introduction} \label{sec:seca}
\newlength{\gnat}
\setlength{\gnat}{24pt} 
\baselineskip=\gnat

Most naturally occurring processes are inherently multivariate in their origination. 
Simultaneous analysis of multiple random variables reveals insights about the relationship between variables. 
This leads to more compelling analysis than marginal consideration of the components alone. 
Multivariate analysis is considerably more complicated than univariate analysis, 
especially when the assumption of multivariate normality does not apply. 
Methods for reducing the complexity of multivariate observations become essential because of the curse of dimensionality.
Independent component analysis (ICA) is a means for finding a suitable representation of multivariate data. 
ICA may also be applied as a dimension-reduction technique, which estimates non-redundant components that are as 
statistically independent as possible.
We propose statistics for measuring and testing mutual independence and 
introduce a novel statistical framework, with minimal prior assumptions, 
for estimation of latent independent sources $\s$ from observations $\y$. 

In statistical analysis, orthogonal components are often used to find suitable representations of multivariate data. 
Principal component analysis (PCA) measures the strength of 
variabilities of orthogonal linear combinations of 
components.
However, higher-order or nonlinear analyses are often needed to adequately approximate complex joint distributions.
Curvilinear component analysis \citep{demartines1997curvilinear} is a nonlinear extension of PCA that preserves 
the proximity between observations in the ${d}$-dimensional input space as the main features are projected onto 
a ${r}$-dimensional $(r < d)$ subspace.
Canonical correlation analysis \citep{hotelling1936relations} generalizes PCA to find linear relationships 
among two sets of variables.
Multidimensional scaling \citep{borg2005modern} measures the dissimilarities between two sets of variables, 
but it typically does not consider higher order relationships. 

To overcome these weaknesses, we consider modeling multivariate random variables with mutually independent components (ICs).  
ICA is a method of unsupervised statistical learning that evolved in computer science research on artificial neural networks. 
\cite{Hyva:Karh:Oja:inde:2001} provide an extensive overview 
including discussion of non-Gaussianity, some algorithms for estimating ICs, and applications in 
blind source separation, feature extraction, compression and redundancy reduction,
medical signal processing (fMRI, ECG, EEG),  clustering, and time series analysis.
Information theory \citep[see][]{hyvarinen1997fast}, the maximum likelihood principal \citep[see][]{Hast:Tibs:Inde:2002},  generalized decorrelation \citep[see][]{cardoso1989, bach2003kernel}, 
and characteristic functions \citep[see][]{Eriksson:2003, Chen:Bick:cons:2005} are four broad methods for ICA estimation. 

Whereas principal components always exist for variables with finite second moments, independent components may not.
In an important contrast with the existing ICA literature, we do not assume the existence of ICs for a given dataset \emph{a priori}.
This distinction makes our approach more general, with much greater applicability. In particular, our estimator is shown to be consistent 
regardless of whether ICs exist.
$U$-statistics of certain Euclidean distances between sample elements are then combined in succession to construct a robust test for the existence of mutually independent components.

We make two more consequential departures from the existing ICA literature. 
First, all assumptions are placed on the observations $\y$ directly, not the latent components $\s$.
This allows direct assessment of every assumption, whereas assumptions made about ICs minimally require that they in fact exist.
Further, because the ICs are latent, any assumptions made about them cannot be verified directly from the observed data. 
In general, an irreconcilable procedure will result when an estimation method requires prior assumptions about ICs.
Specifically, in order to validate such assumptions, estimates of ICs must be obtained; 
however, these estimates of ICs are only reliably obtained if the assumptions are true. 
Second, our measures and tests of mutual independence are based on both necessary and sufficient conditions for mutual independence. 
Those based only on necessary conditions for mutual independence, such as \cite{cardoso1989}, are clearly not robust to all forms of dependence. 
They simply provide no assurance in the identification of mutually independent components and justifiability should be named and categorized as otherwise. 


A linear combination of ICs captures the essential structure of 
multivariate data in many situations, even when other linear projection methods such as PCA, factor analysis, 
or projection pursuit are not effective. 
When ICs exist, one may apply univariate analysis to study or model each component separately. 
Univariate models may then be combined to obtain a multivariate model for the original observations. 
A static linear latent factor model for vector observations $\y$ is given by
\begin{equation}\label{mixing}
\y = \utwi{M}\s,  
\end{equation}
in which $\utwi{M}$ is a constant, nonsingular {\em mixing matrix}, and $\s$ is a random vector.
The goal is to use observations $\y$ to estimate both $\utwi{M}$ and $\s$, such that 
the components of $\s$ are mutually independent, or as close as possible, given a particular dependence measure. 

For computational simplicity, 
let $\bO$ denote an {\it uncorrelating} matrix and let $\z = \bO \y$ denote uncorrelated observations. 
In practice, transformation of a sample estimate for the covariance of $\y$ can be used to approximate $\bO.$ 
The relationship between $\z$ and $\s$ is then 
\begin{equation}\label{Nt}
\s = \utwi{M}^{-1}\y = \utwi{M}^{-1}\utwi{O}^{-1}\z = \utwi{W}\z, 
\end{equation}
in which $\utwi{W} = \utwi{M}^{-1}\utwi{O}^{-1}$ is referred to as the {\em separating} matrix.  
We seek to estimate a separating matrix $\utwi{W}$ that identifies components which are as independent as possible for a particular sample. 

In the setting described above, all ICA methods are executed either symmetrically or sequentially. 
Symmetric methods jointly estimate all components simultaneously, 
whereas sequential algorithms, also referred to as deflationary, estimate the components of $\s$ one by one. 
Motivated by potential computational savings, the deflationary approach has been widely promoted in the machine learning literature. However, estimation uncertainty accumulates at each stage in the succession, and joint estimation will always 
have greater statistical efficiency. For the methodology we propose we briefly compare the speed and accuracy of joint verses sequential estimation. 
%

In Section 2 we introduce our methodology, discuss parameterization and identifiability, 
propose measures for testing mutual independence, propose a versatile inferential framework based on resampling, and state conditions for the strong consistency of 
the proposed estimator. 
In Section 3 we compare the proposed method with popular alternatives in simulation studies,
detail practical implementation and discuss empirical performance measures.
In Section 4 we apply the proposed approach to two real examples and contrast it with PCA.
Concluding remarks are in Section 5 and 
technical proofs follow in the Appendix. 

\section{Methodology}

Let $\Y = \{Y_i : i = 1,\ldots,n\}$ be an iid sample from the joint distribution of a random vector $\y \in \mathbb{R}^d$.
We require $\y$ to obey some standard regularity conditions.
\begin{asm}\label{obs}
The vector random variable $\y \in \mathbb{R}^d$ has a nonsingular, continuous distribution function $F_Y$, with $E(\y) = {0}$ and $\mathrm{E}|\y|^2 < \infty$. 
\end{asm}
\noindent
The fundamental premise in ICA is that $\y$ can be well 
approximated by a linear combination of ICs via Equation (\ref{mixing}).
The existence of ICs will be checked in applications.

\subsection{Parameterization and Identifiability}

Let $\s = (s_{1},\ldots ,s_{d})'$ denote a random vector of ICs. 
Specifically, the univariate components $s_{1},\ldots ,s_{d}$ are mutually independent. 
The first ambiguity associated with Equation (\ref{mixing}) is the scale of the latent variables. 
Without loss of generality, $\s$ is assumed to be standardized such that $\mathrm{E}(s_{k})=0$ 
and $\mathrm{Var}(s_{k})=1$, for $k = 1,\ldots,d$.

For theoretical and practical considerations it is convenient to work with uncorrelated random variables. 
That is, we employ $\z$ in Equation (\ref{Nt}).  
Let $\bm{\Sigma}_{Y} = \mathrm{Cov}(\y)$ denote the covariance matrix of the random variable $\y$, which is assured to exist by Assumption \ref{obs}.  
Let $\bm{\Upsilon}$ be the matrix of eigenvectors and $\bm{\Lambda}$ the diagonal matrix 
of the corresponding eigenvalues of $\bm{\Sigma}_{Y}$, then take 
$\bO$ = $\bm{\Lambda}^{-1/2}\bm{\Upsilon}'.$ 
Without loss of generality, 
we henceforth assume that $\mathrm{Cov}(\z) =\I_d,$ the $d \times d$ identity matrix.
Given the uncorrelated variable $\z$, Equation (\ref{Nt}) implies that 
the separating matrix $\utwi{W}$ is necessarily orthogonal, because 
$\utwi{I}$ = $\mbox{Cov}(\s)$ = $\utwi{W}\mbox{Cov}(\z)\utwi{W}'$ = $\utwi{W}\utwi{W}'$. 
Therefore, $\w$ has $d(d-1)/2$ free elements, instead of $d^2$. 

For $d \ge 2$, let ${\cal O}(d)$ denote the group of all $d \times d$ orthogonal matrices and let ${\cal SO}(d)$ denote the subgroup (rotation group) with determinant equal to 1. Some relevant properties of ${\cal SO}(d)$ are discussed in \cite{MattesonTsay2011}. 
Let ${\xi}_1,\ldots,{\xi}_d$  denote the canonical basis of $\mathbb{R}^d$. Let $\q_{ij}(\psi)$ denote a rotation of all vectors lying in the (${\xi}_i, {\xi}_j$)-plane of $\mathbb{R}^d$ by an angle $\psi$, oriented such that the rotation from ${\xi}_i$ to ${\xi}_j$ is assumed to be positive. Specifically, for $i \ne j$, $\q_{ij}(\psi)$ is a Givens (plane) rotation matrix, that is, the identity matrix $\I_d$ with the $(i,i)$ and $(j,j)$ elements replaced by $\cos(\psi)$, the $(i,j)$ element replaced by $-\sin(\psi)$, and the $(j,i)$ element replaced by $\sin(\psi)$. 

Let ${\theta}$ denote a length $p = d(d-1)/2$ vectorized triangular array of rotation angles, indexed by $\{ i,j : 1\le i < j \le d \}.$
Any rotation $\w \in {\cal SO}(d)$ can be written in the form
\begin{equation*}\label{q1}
\w_{{\theta}} = \q^{(d-1)} \cdots \q^{(1)},
\quad \mbox{in which} \quad
\q^{(k)} =  \q_{k,d}(\theta_{k,d}) \cdots \q_{k,k+1}(\theta_{k,k+1}). 
\end{equation*}
Although such decompositions are not unique, the one given above has an important invariance property. 
Specifically, the $k$th row of $\w_{{\theta}}$ and the $k$th row of the partial product $\q^{(k)} \cdots \q^{(1)}$ coincide. 
Let $\T^{(\ell:k)} = \{\theta_{i,j} : \ell \le i \le k, i < j \le d \},$ then for $\s = \utwi{W}_{\T}\z,$ we observe that 
the $k$th element of $\s$ only varies with the subset of angles in $\T^{(1:k)}.$ 
%
Let
\be\label{Theta}
\Theta = \left\{  \theta_{i,j} : 
\Bigg\{ \begin{array}{ll}
0 \le \theta_{1,j} < 2\pi, &         \\
0 \le \theta_{i,j} <  \pi, & i \ne 1.
\end{array}   \right\}.
\ee
Then, there exists a unique inverse mapping of $\w \in {\cal SO}(d)$ into ${\theta} \in \Theta,$ such that the mapping is assured to be continuous if either all elements on the main-diagonal of $\w$ are positive, or all elements of $\w$ are nonzero \citep[see][]{matteson2008statistical}. 

There are two remaining ambiguities associated with identification of $\m$ and $\s$, the sign and the order of the ICs. Let $\p_{\pm}^{\phantom{'}}$ denote a signed permutation matrix and note that the linear mixing model $\y = \m \s$ is equivalent to 
\begin{equation*}\label{pmix}
\y = \m \p_{\pm}' \p_{\pm}^{\phantom{'}} \s
= (\m \p_{\pm}' ) ( \p_{\pm}^{\phantom{'}} \s),
\end{equation*}
in which $\p_{\pm}^{\phantom{'}} \s$ are new ICs and $\m  \p_{\pm}'$ is the new mixing matrix. 
When identification of ICs up to a signed permutation is sufficient for modeling purposes we may construct an equivalence class and a canonical form for $\w$ to conduct inference \citep[see][]{MattesonTsay2011}. 
In general, the ambiguities in scale, sign and order for ICs must all be taken into account when comparing different estimates;
a metric which is invariant to all three is discussed in Section 3.

\subsection{Measuring Pairwise Multivariate Independence}

Distance covariance $\V(X^{(1)}, X^{(2)})$ is a multivariate measure of independence between random vectors $X^{(1)} \in \mathbb{R}^{d_1}$ and $X^{(2)} \in \mathbb{R}^{d_2}$ of arbitrary dimensions, $d_1$ and $d_2$, for all distributions with finite first absolute moments. 
Let $|\cdot|$ denote Euclidean distance and let
$(\dot X^{(1)}, \dot X^{(2)})$ and $(\ddot X^{(1)}, \ddot X^{(2)})$ denote iid copies of $(X^{(1)}, X^{(2)})$.
Then \cite{szekely2007measuring} show that distance covariance may be defined as 
\bse\label{Ixy}
{\cal I}(X^{(1)}, X^{(2)}) 
&=& E|X^{(1)} - \dot X^{(1)}||X^{(2)} - \dot X^{(2)}| + E|X^{(1)} - \dot X^{(1)}|E|X^{(2)} - \dot X^{(2)}| \\ & & 
- E|X^{(1)} - \dot X^{(1)}||X^{(2)} - \ddot X^{(2)}|  - E|X^{(1)} - \ddot X^{(1)}||X^{(2)} - \dot X^{(2)}|.
\ese

The following properties of $\V$ are the most relevant for ICA:
$0 \le \V(X^{(1)}, X^{(2)})$; 
$\V$ is invariant to the group of orthogonal transformations such that
$\V({a}_1 + b_1 \utwi{C}_1X^{(1)}, {a}_2 + b_2 \utwi{C}_2X^{(1)}) = \sqrt{|b_1||b_2|}\V(X^{(1)}, X^{(2)})$ for all constant vectors ${a}_1,{a}_2$, non-zero scalars $b_1,b_2$, and orthogonal matrices $\utwi{C}_1, \utwi{C}_2$, of conforming dimensions, respectively;
and finally, $\V(X^{(1)}, X^{(2)}) = 0$ if and only if $X^{(1)}$ and $X^{(2)}$ are independent.

Let $\phi_{1}$ and $\phi_{2}$ denote the characteristic functions of $X^{(1)}$ and $X^{(2)}$, respectively, and let $\phi_{1,2}$ denote the joint characteristic function of $X^{(1)}$ and $X^{(2)}$. 
Distance covariance measures the distance between the joint characteristic function and the product of the marginal characteristic functions. It can be applied to test the following hypothesis of independence
$$ H_0: \phi_{1,2}(\bt) = \phi_{1}(\bt_1) \phi_{2}(\bt_2) \quad \mathrm{vs.} \quad H_A:  \phi_{1,2}(\bt)  \ne \phi_{1}(\bt_1) \phi_{2}(\bt_2), \quad \forall \bt_1 \in \mathbb{R}^{d_1}, \bt_2 \in \mathbb{R}^{d_2},$$
in which $\bt' = (\bt_1',\bt_2').$ 
The equality stated in $H_0$ above is both a necessary and sufficient condition for multivariate independence. 

Let $(\X^{(1)}, \X^{(2)}) = \{ (X_i^{(1)}, X_i^{(2)}) : i = 1,\ldots,n \}$ be an iid sample from the joint distribution of vector random variables $X^{(1)} \in \mathbb{R}^{d_1}$ and $X^{(2)} \in \mathbb{R}^{d_2},$ with $E(|X^{(1)}| + |X^{(2)}|) < \infty$. 
%
We define an empirical multivariate independence measure as  
\be\label{IXY}
{\cal I}_n(\X^{(1)}, \X^{(2)}) = T_{1,n}(\X^{(1)}, \X^{(2)}) + T_{2,n}(\X^{(1)}, \X^{(2)}) - T_{3,n}(\X^{(1)}, \X^{(2)}),
\ee
which is a sum of $U$-statistics defined as
\bse
T_{1,n}(\X^{(1)}, \X^{(2)}) &=& {n \choose 2}^{-1} \sum_{i < j} \big|X^{(1)}_i - X^{(1)}_j \big|\big|X^{(2)}_i - X^{(2)}_j\big|, \\
T_{2,n}(\X^{(1)}, \X^{(2)}) &=& \left[ {n \choose 2}^{-1} \sum_{i < j} \big|X^{(1)}_i - X^{(1)}_j\big| \right] \left[ {n \choose 2}^{-1} \sum_{i < j} \big|X^{(2)}_i - X^{(2)}_j\big| \right], \; \mathrm{and} \\
%
%
T_{3,n}(\X^{(1)}, \X^{(2)}) &=& {n \choose 3}^{-1} \sum_{i < j < k} \frac{1}{3}
\Big( \big|X^{(1)}_i - X^{(1)}_j\big|\big|X^{(2)}_i - X^{(2)}_k\big| + \big|X^{(1)}_i - X^{(1)}_k\big|\big|X^{(2)}_i - X^{(2)}_j\big| \\
& & \phantom{{n \choose 3}^{-1} \sum }  + \big|X^{(1)}_i - X^{(1)}_j\big|\big|X^{(2)}_j - X^{(2)}_k\big| + \big|X^{(1)}_j - X^{(1)}_k\big|\big|X^{(2)}_i - X^{(2)}_j\big|  \\
& & \phantom{{n \choose 3}^{-1} \sum }  + \big|X^{(1)}_i - X^{(1)}_k\big|\big|X^{(2)}_j - X^{(2)}_k\big|  + \big|X^{(1)}_j - X^{(1)}_k\big|\big|X^{(2)}_i - X^{(2)}_k\big| \Big), 
\ese
%
%
%
%
%
respectively. For more extensive discussion on distance covariance, and an alternative, asymptotically equivalent, empirical measure based on $V$-statistics, 
see \cite{szekely2009brownian}, 
from which we note $\lim_{n \rightarrow \infty} {\cal I}_n(\X^{(1)}, \X^{(2)})  \stackrel{a.s.}{=} \V(X^{(1)}, X^{(2)}),$ 
as well as convergence in distribution of $n {\cal I}_n(\X^{(1)}, \X^{(2)})$ to a non-degenerate random variable, under $H_0$. 
Additionally, ${\cal I}_n$ is invariant to the same group of orthogonal transformations as ${\cal I}$.

\subsection{Measuring and Testing for Mutual Independence via $U$-Statistics}

To test whether the univariate components of a random vector $\s \in \mathbb{R}^d$ are mutually independent, we propose a statistic based on distance covariance. 
Let $\bt = (t_1,\ldots,t_d)' \in \mathbb{R}^d$.
A necessary and sufficient condition for ${\s}$ to consist of mutually independent components is that 
$\phi_{S}(\bt) = \phi_{s_1}(t_1) \cdots \phi_{s_d}(t_d), \; \forall \bt \in \mathbb{R}^{d},$
in which $\phi_{S}$ is the joint and $\phi_{s_k},$ $k = 1,\ldots,d,$ are 
the marginal characteristic functions of $\s$, respectively. 
Assuming $\s$ has a continuous distribution, 
let $F_{s_k},$ $k = 1,\ldots,d$, denote the continuous univariate marginal distribution functions of $\s.$ 
When applied to the corresponding component of $s_k,$ each function is a probability integral transformation (PIT), $F_{s_k}: \mathbb{R} \rightarrow [0,1]$, defined as $u_{k} = F_{s_k}(s_{k})$.
The marginal distributions for each transformed component $u_{k}$ is Uniform(0,1). 
Further, ${\s}$ consists of mutually independent components if and only if $U = (u_1,\ldots,u_d)'$ does. 

Let ${k^+} = \{ \ell : k < \ell \le d\}$, that is ${k^+}$ denotes the indices $(k+1),\ldots,d,$ and let $\bt_{k^+} = (t_{k+1},\ldots,t_d)'.$ 
We propose simultaneously testing the following joint hypotheses against the stated alternative
  \[
        \begin{array}{ll}
        H_0: & \phi_{u_k, u_{k^+}}(t_{k},\ldots,t_d)   =  \phi_{u_k}(t_k) \phi_{u_{k^+}}(\bt_{k^+}), \quad \forall \bt \in \mathbb{R}^{d}, \quad \mbox{for all} \; k = 1,\ldots,d-1, \\
        H_A: & \phi_{u_k, u_{k^+}}(t_{k},\ldots,t_d)  \ne  \phi_{u_k}(t_k) \phi_{u_{k^+}}(\bt_{k^+}) , \quad \forall \bt \in \mathbb{R}^{d}, \quad  \mbox{for some} \; k = 1,\ldots,d-1.
        \end{array}
        \]
Note that $H_0$ above is both a necessary and sufficient condition for ${\s}$ to consist of mutually independent components, 
since 
\bse
| \phi_{u_1,\ldots,u_d}(t) - \phi_{u_1}(t_1) \cdots \phi_{u_d}(t_d) | 
&\le& \sum_{k = 1}^{d-1} | \phi_{u_k,u_{k^+}}(t_k,t_{k^+})  - \phi_{u_k}(t_k) \phi_{u_{k+}}(t_{k^+}) |,
\ese
$\forall t \in \mathbb{R}^d$, by the triangle inequality, the multiplicative property of absolute value, and the boundedness of characteristic functions. 

Let $\utwi{S} = \{S_i : i = 1,\ldots,n \}$ be an iid sample from the joint distribution of the vector random variable $\s \in \mathbb{R}^{d}.$ 
Let $\utwi{S}_1, \ldots, \utwi{S}_d$ be a partition of the elements of $\utwi{S}$ into $d$ univariate components.  
In practice, the marginal distribution functions of $\s$ are unknown, so we replace each PIT with its empirical counterpart. 
Specifically, for each component of $\utwi{S},$ we replace each observation with its normalized marginal rank.
That is, each component-wise transformation $\widehat{\utwi{U}}_k$ is defined as $\hat{{u}}_{i,k} = \frac{1}{n}\mathrm{rank} \{S_{i,k} : S_{i,k} \in \utwi{S}_k \},$  for each $k = 1,\ldots,d.$
Finally, we define a test statistic for mutual independence as
\begin{equation}\label{test}
{\cal U}_n(\utwi{S}) =  n \sum_{k=1}^{d-1} \,  \V_n(\widehat{\utwi{U}}_k, \widehat{\utwi{U}}_{k^+}).
\end{equation}

For $d = 2$, ${\cal U}_n(\utwi{S})$ is asymptotically distribution free and its 
asymptotic distribution can be derived from Theorem 5 of \cite{szekely2009brownian} 
and the Glivenko-Cantelli theorem. 
For the more general case, the distribution of ${\cal U}_n(\utwi{S})$ depends on the distribution of $\s,$
and in practice we implement a permutation test.
%
%
The null hypothesis of mutual independence is rejected for a large value of ${\cal U}_n(\utwi{S}).$
Similar to \cite{szekely2007measuring}, we note that if any subsets of $\s$ are dependent, then 
${\cal U}_n(\utwi{S}) \rightarrow \infty$ in probability, as $n \rightarrow \infty.$ Hence, the proposed test of mutual independence is also
statistically consistent against all types of dependence. 

\subsection{Estimation of Independent Components via $U$-Statistics}

Let $\utwi{Y}$ be an iid sample from the joint distribution of the continuous vector random variable $\y.$
In practice, $\utwi{Y}$ is usually replaced by a centered version $\widehat{\utwi{Y}},$ in which the sample mean vector is subtracted from each observation. 
Recall that an uncorrelated variable $\z$ can be defined as $\z = \bO \y,$ in which $\bO$ denotes an uncorrelating matrix.
In practice, $Cov(Y)$ is unknown, however, under Assumption \ref{obs}, the sample covariance provides a consistent estimate.
That is, $\widehat{Cov}_n({\utwi{Y}}) \stackrel{a.s.}{\longrightarrow} Cov(Y),$ as $n \rightarrow \infty.$
Using the sample covariance we can approximate the uncorrelating matrix as $\widehat{\bO}_n = \widehat{Cov}_n({\utwi{Y}})^{-1/2}$, 
then define approximately uncorrelated observations as $\widehat{\utwi{Z}}_n = {\utwi{Y}} \widehat{\bO}_n'.$
This is done such that $\widehat{Cov}_n(\widehat{\utwi{Z}}_n) = \I_d, \; \forall n,$ and  
${Cov}(\widehat{\utwi{Z}}_n) \stackrel{a.s.}{\longrightarrow} \I_d,$ as $n \rightarrow \infty.$

To simplify notation, we omit the steps described above, and let 
 $\utwi{Z},$ an uncorrelated, mean zero, unit variance, iid sample, be given. 
We begin by estimating $\w_{{\theta}}$ via ${\theta}$. 
Define
$\s({\theta}) = \w_{{\theta}}\z$, 
$\utwi{S}({\theta}) = \utwi{Z} \w_{{\theta}}'$, and let $\utwi{S}_k({\theta})$ denote the $k$th component 
of $\utwi{S}({\theta}).$ 
Recall that, by the construction of $\w_{{\theta}},$ each $\utwi{S}_k({\theta})$ only varies with the subset of angles in $\T^{(1:k)},$ in which $\T^{(\ell:k)} = \{\theta_{i,j} : \ell \le i \le k, i < j \le d \}$,
and it is invariant to the complementary subset. 

Recall ${k^+} = \{ \ell : k < \ell \le d\}.$
To find a sample $\utwi{S}({\theta})$ which has mutually independent components, we define an objective function as
\begin{eqnarray}\label{J_n2}
{\cal J}_n({\theta}) = \sum_{k=1}^{d-1} \, \V_n(\utwi{S}_k(\T), \utwi{S}_{k^+}(\T)), 
\end{eqnarray}
and we define the distance covariance ICA estimator (dCovICA) as 
$\widehat{{\theta}}_n = \argmin_{{\theta}} {\cal J}_n({\theta}).$
Given an estimate of ${\theta}$, the separating matrix is estimated as 
$\w_{\widehat{{\theta}}_n}$ and the estimated ICs $\widehat{\utwi{S}}$ are given by the components of $\utwi{S}(\widehat{{\theta}}_n) = \utwi{Z} \w_{{\widehat{{\theta}}_n}}'$.

The objective function in Equation (\ref{J_n2}) has $d(d-1)/2$ parameters which can be estimated jointly.
Alternatively, estimation may be preformed conditionally in a sequence of $d-1$ minimization problems; 
the first will have $d-1$ parameters, the second $d-2,$ continuing as such until the last, which will have one parameter.
This follows by the orthogonal invariance property of $\V_n$ stated in Section 2.2.
Specifically, 
let $\widehat{\T}^{(1:1)} = \argmin_{\T^{(1:1)}} \V_n(\utwi{S}_1(\T), \utwi{S}_{1^+}(\T))$,
in which the elements ${\T}^{(2:d)}$ are fixed, but arbitrary.
Now, for $k = 2,\ldots,(d-1)$, given $\widehat{\T}^{(1:(k-1))}$, let
\be\label{J_n3}  
\widehat{{\theta}}_n^{(k:k)} = \{\widehat{\theta}_{k,\ell} : k < \ell \le d \} = \argmin_{{{\theta}}^{(k:k)}} \V_n (\utwi{S}_k(\T), \utwi{S}_{k^+}(\T)), 
\ee 
in which ${\T}^{(1:(k-1))}$ are fixed at $\widehat{\T}_n^{(1:(k-1))}$ and all elements in ${\T}^{((k+1):d)}$ are fixed, but arbitrary. 
Hence, the sequence of estimates from Equation (\ref{J_n3}), for $k = 1,\ldots,(d-1),$ exactly coincide with the joint estimate 
$\widehat{{\theta}}_n = \argmin_{{\theta}} {\cal J}_n({\theta}).$ 
When the components are estimated in this sequential manner the later component estimates are restricted to lie within the subspace orthogonal to the span of the earlier estimates, resulting in a tradeoff between computational complexity and statistical efficiency.

\subsubsection{An Alternative Estimator}

In general, distance covariance depends on the marginal distributions of the inputs. 
As described in Section 2.3 \citep[also see][]{remillard2009discussion}, 
for continuous random variables, this dependency can be removed by applying the PIT component-wise. 
As before, the marginal distributions functions $F_{s_k}$ are unknown in practice, and the PIT must be approximated. 

Our asymptotic results and our optimization algorithms rely explicitly on our objective function varying continuously in its arguments. 
This means that approximating $F_{s_k}$ using the empirical cumulative distribution functions (CDF) will not be sufficient because it is a step function. 
Simply interpolating the empirical CDF between the steps is also insufficient.
Instead, we require an estimate of $F_{s_k}$ to depend on the location of all the observations $\{s_{i,k} : i=1,\ldots,n\}$, not just their relative location.  
To assure this, we propose applying kernel smoothing to approximate the CDF of each $F_{s_k}$ with a continuous function. 
Let 
\begin{eqnarray}\label{J_n4}
\widetilde{F}_{s_k,n,\tilde{h}_n}(s) = \sum_{i=1}^n   G\left( \frac{s_{i,k} - s}{\tilde{h}_n} \right)
\end{eqnarray}
in which $G$ is the integral of a density kernel
and $\tilde{h}_n$ is a data-dependent bandwidth.
In applications we let $G$ be the Gaussian CDF. The choice of bandwidth is discussed below.

Given $\utwi{Z},$ for $\utwi{S}({\theta}) = \utwi{Z} \w_{{\theta}}'$, we define $\widetilde{\utwi{U}}_k(\T),$ as a continuous function of $\theta,$ such that 
$\tilde{u}_{i,k}(\T) = \widetilde{F}_{s_k(\theta),n,\tilde{h}_n} [s_{i,k}(\T) ],$ for each $k = 1,\ldots,d.$ 
Now, as an alternative objective function, we consider
\begin{eqnarray}\label{J_n5}
\widetilde{{\cal J}}_n({\theta}) = \sum_{k=1}^{d-1} \, \V_n(\widetilde{\utwi{U}}_k(\T), \widetilde{\utwi{U}}_{k^+}(\T)) .
\end{eqnarray}
Finally, we define this PIT and distance covariance based ICA estimator (PITdCovICA) as 
$\widetilde{{\theta}}_n = \argmin_{{\theta}} \widetilde{{\cal J}}_n({\theta}).$ 
Similar to the dCovICA estimator, estimation may also be preformed conditionally in a sequence because invariance to orthogonal transformations is preserved despite the PIT.
Many alternative smoothing methods are available for estimating $F_{s_k}$, but computationally fast methods, such as our proposal, should be strictly preferred since the approximation needs to be updated continuously within any optimization algorithm applied to Equation (\ref{J_n5}). 
The PITdCovICA estimator is computationally more demanding, but it is even more robust to extreme observations 
and it remains invariant to component-wise monotone transformations of the observations $\Y$.
Practical implementation of both estimators is discussed in Section 3.

\subsection{Asymptotic Properties of the Proposed Estimators}

Asymptotic results for the proposed estimators require some basic assumptions about how the observations are transformed and the parameter space. By Assumption \ref{obs} and Slutsky's Theorem, 
without loss of generality, assume throughout this section that $\mbox{E}(Y) = \mathbf{0}$ and $\mathrm{Cov}(Y) = \I_d$, such that $\z = \y$ and $\utwi{Z} = \utwi{Y}.$ 
Let ${{U}}(\T)$ and $\utwi{{U}}(\T)$ be defined as a function of $\theta,$ such that 
${U}_{k}(\T) = F_{s_k}[s_{k}(\T) ],$ and ${u}_{i,k}(\T) = F_{s_k}[s_{i,k}(\T) ],$ for each $k = 1,\ldots,d.$ 
Define the population counterpart of Equation (\ref{J_n5}) as 
\begin{eqnarray}\label{J}
\widetilde{{\cal J}}({\theta}) = \sum_{k=1}^{d-1} \, \V({U}_k(\T), {U}_{k^+}(\T)), 
\end{eqnarray} 
and let $\overline{\Theta}$ denote a sufficiently large compact subset of the space ${\Theta}$ defined by Equation (\ref{Theta}). 
%
To establish uniform a.s.\ convergence of $\widetilde{{\cal J}}_n({\theta})$ to $\widetilde{{\cal J}}({\theta})$ we require 
\begin{eqnarray}\label{unif}
\sup_{y \in \mathbb{R}} 
|\widetilde{F}_{y_k,n,\tilde{h}_n}(y) - {F}_{y_k}(y) | \stackrel{a.s.}{\rightarrow} 0, \;\; \mbox{as} \;\; n \rightarrow \infty,
\end{eqnarray} 
for each component of $Y$.
The Glivenko-Cantelli theorem does not hold for the standard kernel distribution estimators as defined in 
Equation (\ref{J_n4}) with $h_n$ replacing $\tilde{h}_n$.
That is, convergence cannot be established uniformly over all $F \in {\cal F}$, the class of all continuous 
distribution functions \citep{zielinski2007kernel}.
To establish uniform in bandwidth consistency for all $F \in {\cal F}$, a data-driven bandwidth $\tilde{h}_n$ is required. 

\begin{asm}\label{kernel} 
The bandwidth $\tilde{h}_n$ is a measurable function of $\{y_{i,k} : i = 1,\ldots,n \},$ such that $\tilde{h}_n \stackrel{a.s.}{\rightarrow} 0$ as $n \rightarrow \infty,$
and the kernel function $G$ is Lipschitz continuous. 
\end{asm}
Note that these assumptions are made on the kernel distribution estimators, not on the observations.
Equation (\ref{unif}) holds under Assumptions \ref{obs} and \ref{kernel} \citep[see][Corollary 1]{chacon2010note}.


\begin{thm}\label{theta.as}
If Assumptions \ref{obs} and \ref{kernel} hold, if there exists a unique minimizer ${\theta}_0 \in \overline{\Theta}$ of Equation (\ref{J}), and if $\w_{\T_0}$ satisfies the conditions for a unique continuous inverse to exist, then
%
$\widetilde{{\theta}}_n
   \stackrel{\mbox{a.s.}}{\longrightarrow} {\theta}_0,$ as $n {\rightarrow} \infty.$ 
\end{thm}

Convergence of the PITdCovICA estimator is established on equivalence classes; a proof is given in the Appendix. 
Under the same conditions, proof that the dCovICA estimator, based on Equation (\ref{J_n2}), converges a.s.\ 
follows from similar arguments. 


\subsection{Inference Based on Resampling}

Although the minimizers $\widehat{\T}_n$ and $\widetilde{\T}_n$ of Equations (\ref{J_n2}) and (\ref{J_n5}), respectively, always exist,
an important question for all ICA methods is whether the ICs exist or not. 
To evaluate this issue statistically, we construct a test of the null hypothesis
\bse
H_0 : \Y = \utwi{S} \m',
\ese
in which $\m$ is nonsingular and $\utwi{S}_1, \ldots, \utwi{S}_d$ are mutually independent vectors, each of which is a sequence of iid random variables with mean 0 and variance 1.
Under the assumption of linear mixing, the null hypothesis above is a sufficient but not a necessary condition for the existence of ICs. 
Each sequence is only required to consist of identically distributed random variables, but independent sequences are required to construct an estimate of the null distribution via resampling.

Since $\m$ is unknown in practice, we do not observe $\utwi{S}$ directly, and since the limiting distribution of ${\cal U}_n(\widehat{\utwi{S}})$ is 
different than that of ${\cal U}_n(\utwi{S}),$ we define a resampling based procedure below.
This allows us to assess how large is sufficiently large to reject $H_0$ above. If $H_0$ fails to be rejected, we may also construct confidence sets for the mixing matrix $\m,$ and even the ICs $\utwi{S},$ based on the same resampling scheme.
Define $\widehat{\m}_n = {\widehat{\bO}}_n^{-1} \w_{{\T}_n}^{-1}$ as the estimated mixing matrix, in which ${\widehat{\bO}_n}$ is the estimated uncorrelating matrix, and ${\T}_n$ is either the dCovICA estimator $\widehat{\T}_n$ or the PITdCovICA estimator $\widetilde{\T}_n$, as defined in Section 2.4.
The proposed resampling scheme consists of the following two steps.


\begin{enumerate}
\item[(i)]	For $k = 1,\ldots,d,$ 
jointly sample the entire sequence $\utwi{S}_{k}^{*} = (s_{1,k}^{*},\ldots,s_{n,k}^{*})'$ by randomly permuting the $n$ elements of $\widehat{\utwi{S}}_{k}.$

\item[(ii)]	Let $\Y^{*} = {\utwi{S}}^* \widehat{\m}_n',$ and randomly generate a $d \times d$ signed permutation matrix $\p_{\pm}^*.$

\end{enumerate}

\subsubsection{A Test for the Existence of ICs}
First the observed sample $\Y$ is replaced by $\Y^{*}.$
Then, given $\Y^{*},$ the resampled estimator ${\m}^*$ is calculated via the same procedure used to calculate $\widehat{\m}_n.$
We define the resampled ICs estimator as $\widehat{\utwi{S}}^* = \Y^{*} {{\m}^{*'}}^{-1}.$
Let ${\cal U}_n^*(\widehat{\utwi{S}}) = {\cal U}_n({\widehat{\utwi{S}}}^* \p_{\pm}^*).$ 
Under $H_0,$ the limiting distribution of ${\cal U}_n({\utwi{S}})$ is invariant with respect to the ordering of the components of $\utwi{S}.$
For small samples, multiplication by $\p_{\pm}^*$ is recommended to eliminate any possible order dependence from the statistic's distribution. 

Note that the resampled observations $\Y^{*}$ are generated following the model given in $H_0,$ in which the components of ${\utwi{Y}}^* \widehat{\m}_n'^{-1}$ are genuine ICs. 
Hence, under $H_0,$ and conditional on the original observations $\Y,$ the empirical distribution of ${\cal U}_n^*(\widehat{\utwi{S}})$
provides an approximation for the distribution of ${\cal U}_n(\widehat{\utwi{S}}).$ 
Therefore, we repeat the above resampling $N,$ a large integer, times.
Then we reject $H_0$ if ${\cal U}_n(\widehat{\utwi{S}})$ is greater than the ($N\alpha$)th largest value of the ${\cal U}_n^*(\widehat{\utwi{S}}),$ in which $\alpha \in (0,1)$ is the size of the test. 
This test for the existence of ICs accounts for the uncertainty in estimating ICs given approximately uncorrelated observations $\widehat{\utwi{Z}}_n,$ as well as 
the uncertainty in estimating $\widehat{\utwi{Z}}_n.$
This procedure is independent of the estimation method, hence it may be used with any ICA estimation technique.

\subsubsection{Confidence Sets for $\m$}

Let $D(\cdot,\cdot)$ be a suitable metric for comparing two mixing matrices; a specific metric with pertinent invariance properties is defined in Section 3 below. 
A resampling-based approximation for a $1-\alpha$ confidence set of the mixing matrix $\m$ may then be constructed as
\bse
\{ \m : D(\m, \widehat{\m}_n) \le c_{\alpha},  \m \; \mathrm{non}\mathrm{-}\mathrm{singular} \},
\ese
in which $c_{\alpha}$ is the ($N\alpha$)th largest value of $D(\m^*,\widehat{\m}_n)$ obtained in $N$ replications of the resampling scheme.
A confidence set for the separating matrix $\w$ may similarly be defined.

\section{Simulation Performance and Practical Implementation}

In this section we compare the proposed estimation methods with popular alternatives in simulation studies.
We also detail practical implementation and discuss empirical performance measures for ICA. 

\begin{landscape}
\begin{figure}
\centering
\includegraphics[width = 9in, height = 6 in]{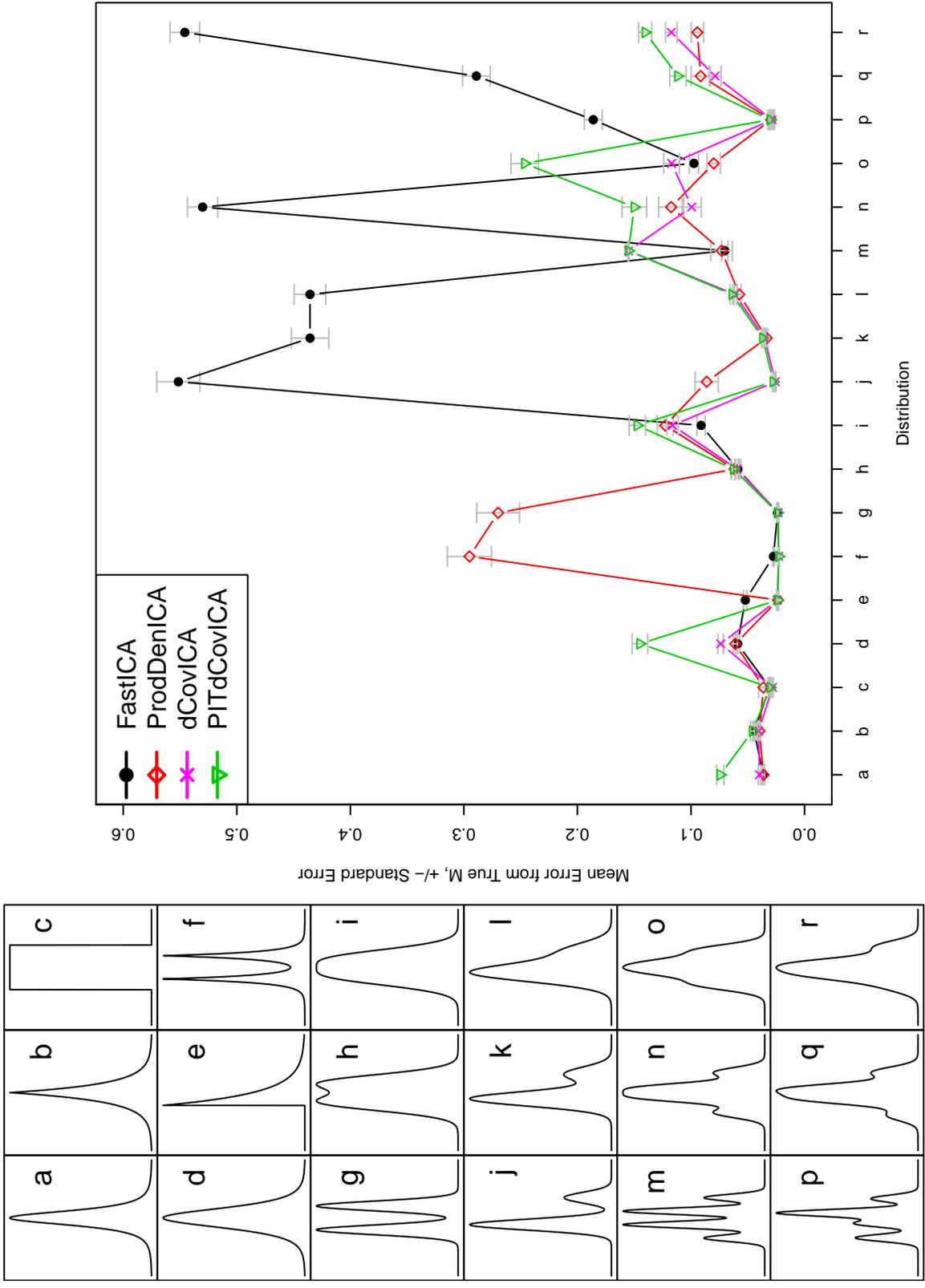}
\caption{
\baselineskip=10pt
The left panel shows 18 distributions used for comparisons. These include the Student-$t$, uniform, exponential, mixtures of exponentials, symmetric and asymmetric Gaussian mixtures. 
The right panel show the mean error distance, Equation (\ref{error}), for each method and each distribution, based on $N = 1000$ simulations in $\mathbb{R}^2$ with sample size $n = 1000$ for each distribution. Vertical bars denote approximate standard errors.
} 
\label{AllSim}
\end{figure}
\end{landscape}

We evaluate the performance of the proposed dCovICA and \mbox{PITdCovICA} estimators by performing 
simulations similar to \cite{bach2003kernel} and \cite{Hast:Tibs:Inde:2002}.
The left panel in Figure \ref{AllSim} shows the 18 distributions used. These include the Student-$t$, 
uniform, exponential, mixtures of exponentials, as well as symmetric and asymmetric Gaussian mixtures. 
For each of these distributions, we simulate ICs $\utwi{S}_0$ with length $n = 1000$ and 
a random mixing matrix $\m_0 \in \mathbb{R}^{2\times2}$ with condition number between $1$ and $2$ using  
the {\tt R} \citep{R} package {\tt ProDenICA} \citep{ProDenICA}. Observations are then defined as $\Y_0 =  \utwi{S}_0 \m_0'.$
We compare empirical performance of the proposed estimators with the FastICA estimator using the negentropy criterion \citep{hyvarinen1997fast} and the ProDenICA estimator using a tilted Gaussian density \citep{Hast:Tibs:Inde:2002}.

The simulated observations are centered by their sample mean, then pre-whitened using the standardized scores from PCA. 
In practice, ICA typically requires minimization of a non-linear, locally convex objective function. This is performed using iterative algorithms, any of which requires initialization.
To find a suitable initialization, we perform Latin hypercube sampling uniformly over the space ${\Theta}$ defined in Equation (\ref{Theta}) to obtain 1000 parameter values. We then evaluate the objective function at each value and record which minimizes the objective function. This is used to initialize the corresponding algorithms. 
We recommend that the number of parameter values considered should grow with the dimension. 

Each method returns an estimate for the mixing matrix. To jointly measure the uncertainty associated with pre-whitening and estimating ICs, we use the metric proposed by \cite{ilmonen2010new} to measure the error between an estimate $\widehat{\m}$ and the known parameter $\m_0.$
It is defined as
\be\label{error}
D(\m_0, \widehat{\m}) = \frac{1}{\sqrt{d-1}} \inf_{{C} \in {\cal C}} || \utwi{C} \widehat{\m}^{-1} \m_0 - \I_d||_F,
\ee
in which $|| \cdot ||_F$ denotes the Frobenius norm. 
Let ${\cal M}$ be the set of $d \times d$ nonsingular matrices. 
Let $\p_{\pm}$ be a signed permutation and let $\utwi{B}$ be a diagonal matrix with positive diagonal elements, both $d \times d$.
The infimum above is taken such that the metric $D$ is invariant with respect to the three ambiguities associated with ICA by defining 
\bse
{\cal C} = \{  \utwi{C} \in {\cal M} : \utwi{C} =  \p_{\pm}  \utwi{B} \hbox{ for some }   \p_{\pm} \hbox{ and } \utwi{B}  \}.
\ese
A function for computing $D$ is available in the {\tt R} package {\tt JADE} \citep{JADE}.

The right panel of Figure \ref{AllSim} shows the mean error for each method and each distribution, based on $N = 1000$ simulations for each distribution, with vertical bars for standard errors. 
The dCovICA and \mbox{PITdCovICA} results are competitive with FastICA and ProDenICA in all situations.
FastICA is dominated for most of the mixture distributions. ProDenICA is less accurate for several of the multimodal distributions.  
For $n = 1000,$ we see that dCovICA outperforms \mbox{PITdCovICA} in some cases as well. 

Let $\mathrm{IQR}_n$ denote the interquartile range, then, following \cite{silverman1986density}, 
the bandwidth of a Gaussian kernel distribution estimator $\hat{h}_n$ is chosen as 
\bse
0.9 \min \left\{ \widehat{\mathrm{sd}}_n(\bS_k), \frac{\mathrm{IQR}_n(\bS_k)}{1.34} \right\} n^{-1/5}.
\ese
%
If Assumption \ref{obs} holds, then Assumption \ref{kernel} is also satisfied by this bandwidth choice since $\hat{h}_n \stackrel{a.s.}{\rightarrow}  0$ as $n \rightarrow \infty.$
Other bandwidth choices have been proposed; \cite{scott1992multivariate} uses a factor of 1.06.
To investigate the finite sample effect the bandwidth choice has on the PITdCovICA estimator, we repeated the previous simulation adjusting the Silverman rule bandwidth by a scale factor of 0.25, 0.5, 1, 1.5, and 2.
The difference in mean error for the PITdCovICA method with these bandwidth adjustments was much smaller then  
the size of the standard errors, so we conclude there is no significant difference between these bandwidths in this simulation. 

Finally, with $n= 1000$, we also ran $N = 1000$ simulations in $\mathbb{R}^4$, $\mathbb{R}^6$, and $\mathbb{R}^8$ by randomly selecting 4, 6, or 8 of the 18 distributions, respectively, for each iteration and generating $\m_0$ as above. The results are shown in Table \ref{sim}, including mean computation times. 
FastICA was much faster on average, but its mean error was about twice as large as the others. 
ProDenICA was slightly faster than the proposed methods on average.
We included both joint and sequential estimation of the dCovICA and PITdCovICA estimators for further comparison. 
Joint estimation of the PITdCovICA estimator had the smallest mean error, but was also the slowest. 
The mean error for sequential estimation increased more quickly with the dimension relative to the corresponding joint estimators.

\section{Application}
In this section we illustrate and discuss application of our methodology to two real examples. 
Throughout this section the PITdCovICA estimator is calculated using joint estimation.
We use the Gaussian kernel, with Silverman's rule to choose the bandwidth, and 1000 starting values, as outlined in Section 3.

\subsection{U.S. Crime Rate}

The Freedman data \citep{freedman1975crowding}, from the 
U.S. Census Bureau,
reports crime rates in U.S. metropolitan areas with 1968 populations of 250,000 or more. 
The data are available in the {\tt R} package {\tt car} \citep{car}.
We consider four variables: 
the logarithm of population (1968 total, in thousands);
nonwhite (percent nonwhite population, 1960); 
density (population per square mile, 1968); 
and crime (crime rate per 100,000, 1969). 
The main interest is identifying the primary determinants of the crime rate. 

To simplify our analysis we first remove the 10 observations with missing values and analyze $n = 100$ cities with complete data. 
Next, the sample mean was subtracted from each observation. 
Finally, each of the four marginal variables is divided by its sample standard deviation $(0.79, 10.08, 1441.95, 983.58)'$ to simplify parameter interpretation. 
Now, we test whether these standardized observations $\widehat{\Y}$ are ICs using the statistic from Equation (\ref{test}).
The test statistic is 
${\cal U}_n(\widehat{\Y}) = 2.52,$ with $p$-value $\approx 0,$
indicating significant dependence. 
Next PCA was applied to obtain approximately uncorrelated components $\widehat{\utwi{Z}}$. 
The ICs test statistic for these standardized PC scores is  
${\cal U}_n(\widehat{\utwi{Z}}) = 1.59,$ with $p$-value $\approx 0,$
hence the PCs are not ICs. 
Finally, ICs $\widehat{\utwi{S}}$ are estimated using the PITdCovICA method.
The ICs test statistic is 
${\cal U}_n(\widehat{\utwi{S}}) = 0.04,$ with $p$-value $\approx 0.37,$
hence, we conclude that ICs do exist for this dataset. 
The estimated mixing matrix and its inverse are shown Table \ref{fig41}.
 %
%
We see that crime is a weighted average of $\hat{s}_1, \hat{s}_3,$ and $\hat{s}_4,$ with loadings 
$0.76$, $0.51$, and $-0.38$, respectively. 

\begin{figure}
\centering
\includegraphics[width = 6.5 in, height = 6.5 in]{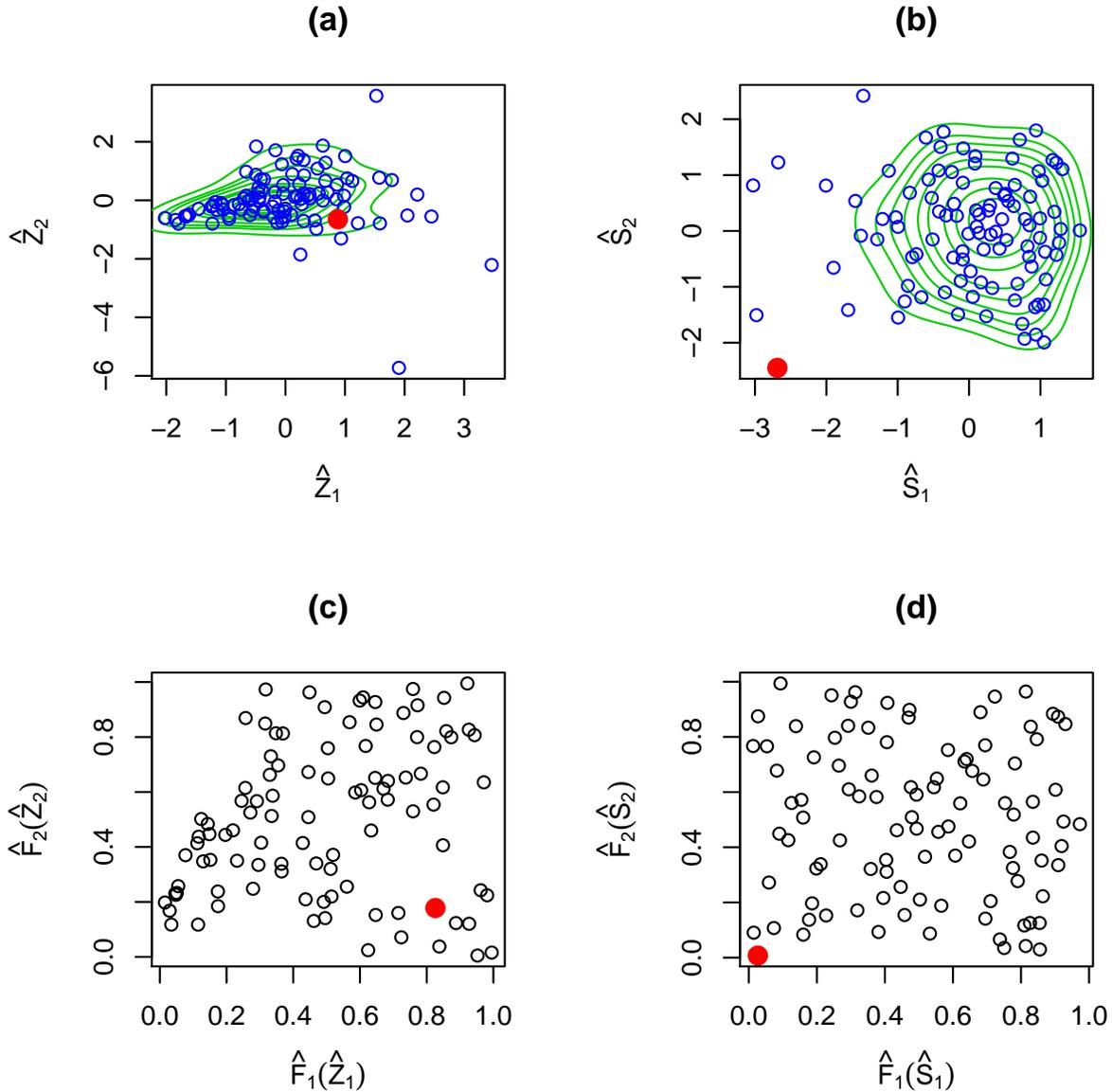}
\vspace{-20pt}
\caption{
\baselineskip=0pt 
 {
The Freedman data based on crime rates in US metropolitan areas with 1968 populations of 250,000 or more. 
We consider four variables: the logarithm of population (total 1968, in thousands), nonwhite (percent nonwhite population, 1960), density (population per square mile, 1968), crime (crime rate per 100,000, 1969). 
(a) first two principal component scores; 
(b) first two estimated independent components;
(c) first two principal component scores and 
(d) first two estimated independent components, each after taking the probability integral transformation 
defined by Equation (\ref{J_n4}).
Estimated contour lines have been drawn for each decile and Philadelphia is indicated on the plot as a larger solid point.
}
}
\label{ICAvPCA}
\end{figure}

\cite{szekely2009brownian} use this dataset to illustrate a jackknife procedure, based on distance covariance, to identify possible influential observations. Their analysis suggests that Philadelphia is an unusual observation. 
The PCs are ordered by the proportion of variability they explain in the observations. 
The first two PCs are shown in Figure \ref{ICAvPCA}(a). Estimated contour lines have been drawn for each decile, and Philadelphia is indicated on the plot as a larger solid point. 
PCA does not identify Philadelphia as an unusual observation, by this plot, or in plots of other pairs of PCs.

The estimated ICs do not have a natural ordering, but $\hat{s}_1$ and $\hat{s}_2$ explain the largest proportion of variability in the observations. 
They are shown in Figure \ref{ICAvPCA}(b), with features similar to \ref{ICAvPCA}(a) included. 
The point corresponding to Philadelphia simultaneously takes large negative values on both $\hat{s}_1$ and $\hat{s}_2$.
%
From Table \ref{fig41} we see that $\hat{s}_1$ has a negative coefficient for population, but positive for the others,
while $\hat{s}_2$ has a positive coefficient for crime, but negative for the others.
This corresponds directly with Philadelphia's relatively low crime rate and its relatively high population level, during this time period.
Figure \ref{ICAvPCA}(c) and Figure \ref{ICAvPCA}(d) show the same observations after taking the empirical PIT component-wise.
A clear trend is visible in Figure \ref{ICAvPCA}(c) confirming rejection of the ICs test  for the PCs,
whereas points in \ref{ICAvPCA}(d) appear uniformly distributed within the unit square.

\subsection{U.S. Unemployment Rate}

To further illustrate the proposed approach we consider analysis of statewide, seasonally adjusted monthly unemployment rates
from January 1976 through August 2010. 
We will focus on six states: CA, FL, IL, MI, OH, and WI. 
The data is available from the U.S. Department of Labor at 
\url{http://Data.bls.gov/cgi-bin/surveymost?la},
and also from {\it FRED} of the Federal Reserve Bank of St.\ Louis 
\url{http://research.stlouisfed.org/fred}.

\begin{landscape}
\begin{figure}
\centering
\includegraphics[width = 9in, height = 6 in]{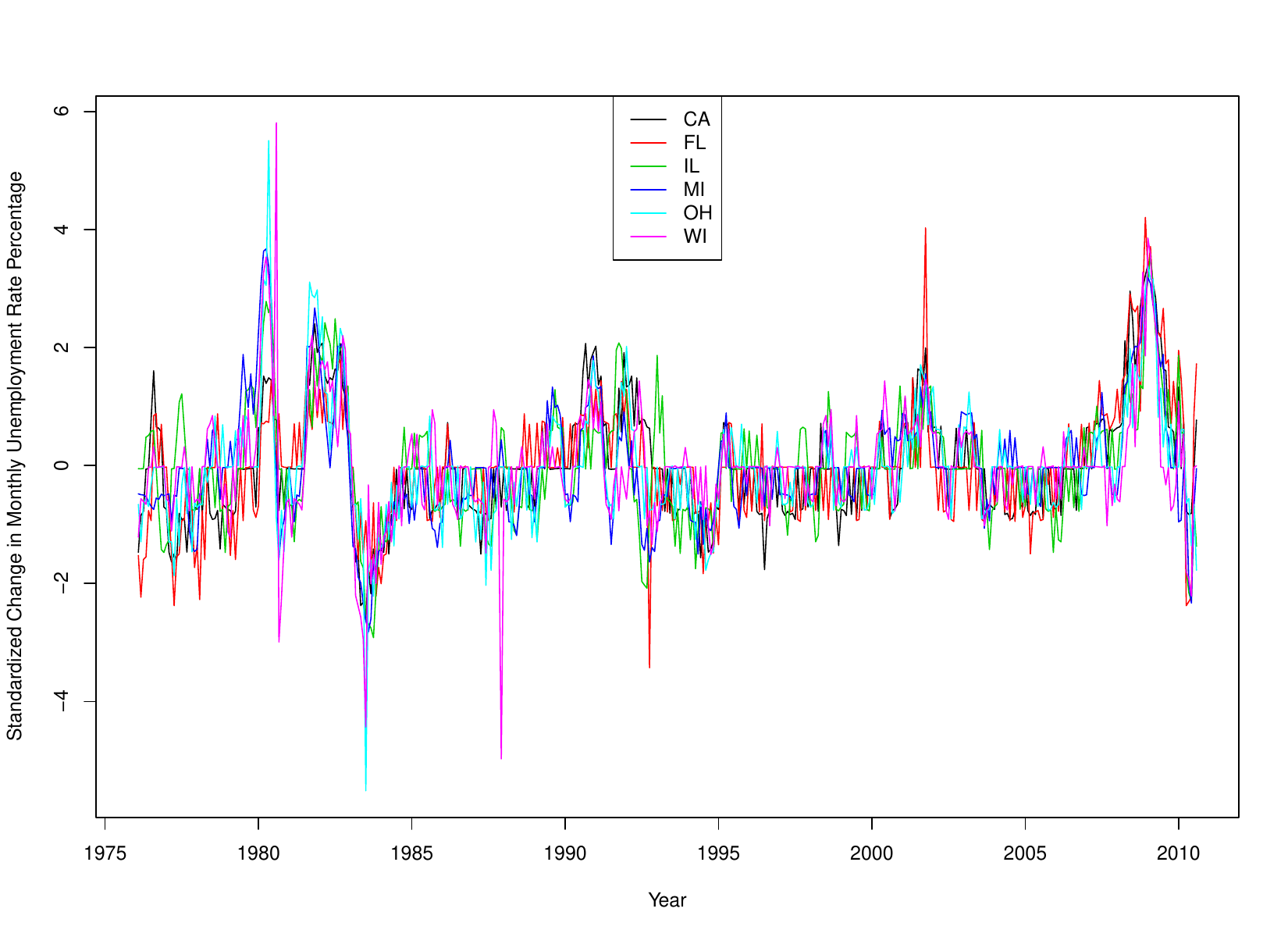} 
\caption{
\baselineskip=10pt
Standardized Change in Monthly Unemployment Rate Percentage for 
California, Florida, Illinois, Michigan, Ohio, and Wisconsin.
This vector series appears stationary, but exhibits serial dependence. 
}
\label{UnempRateDiffStd}
\end{figure}
\end{landscape}

To begin the analysis we difference each series to remove the observed nonstationarity in mean; no trend is present after differencing.  
Next, we scale the observations by the reciprocal of their sample monthly standard deviations 
to remove the observed heteroskedasticity in each series. 
Let $\widehat{\Y}$ denote these standardized observations; they are shown in Figure \ref{UnempRateDiffStd}.
Assumption \ref{obs} also requires the observations to be independent, in this case, over time.  
Let $\y_i$ denote a length $d$ random vector observation occurring at time $i$ and let $\y_{(i-1):(i-m)}' = (\y_{i-1}', \ldots, \y_{i-m}')$
denote a length $dm$ vector containing $m$ observations occurring at times $i-m,\ldots,i-1,$ respectively.
We can use distance covariance to simultaneously measure serial dependence by testing whether or not $\V(\y_i, \y_{(i-1):(i-m)}) = 0.$
Equivalently, we may preform a PIT and base the test on transformed variables $\utwi{U},$ as in Section 2.3.
%

For a $d$ dimensional process $\Y$, with length $n$, we define a joint $m$-lag test statistic as
\be\label{Qtest}
{\cal Q}_d(\Y, m) = (n-m) \V_n \left[ \widehat{\utwi{U}}^{(1+m):n}, \left(\widehat{\utwi{U}}^{(m):(n-1)}, \ldots, \widehat{\utwi{U}}^{1:(n-m)}\right) \right], 
\ee
in which $\widehat{\utwi{U}}$ is the component-wise marginal ranks of $\Y$ and the superscripts denote the observation indices included in each term.
Let $\phi_{u_i}$ denote the joint characteristic function for the transformed variable $U$ at time $i$. The hypothesis we are testing is 
$H_0: \phi_{u_i, u_{i-1}, \ldots, u_{i-m}}(t)  =  \phi_{u_i}(t_1) \phi_{u_{i-1}, \ldots, u_{i-m}}(t_2,\ldots,t_{d+1}),$  $\forall t \in \mathbb{R}^{d(m+1)}$ and $\forall i \in \mathbb{N}$. 
Under the assumption of stationarity, this is equivalent to 
$H_0: \phi_{u_i, u_{i-1}, \ldots, u_{i-m}}(t)  =  \phi_{u_i}(t_1) \phi_{u_{i-1}}(t_2) \cdots \phi_{u_{i-m}}(t_{d+1}),$  $\forall t \in \mathbb{R}^{d(m+1)}$ and $\forall i \in \mathbb{N}$, 
that is, mutual independence between any $m+1$ neighboring observtions. 

\begin{lem}\label{dist} 
Suppose $\Y = \{Y_i : i = 1,\ldots,n\}$ are identically distributed and have a continuous distribution. If they are mutually independent, then for any $m$, 
\bse 
{\cal Q}_d(\Y, m) \stackrel{D}{\longrightarrow} Q, \;\; as \;\; n \rightarrow \infty,
\ese 
in which $Q$ is a non-degenerate random variable. 
%
%
\end{lem}
\noindent
The definition of $Q$ and its distribution can be derived from Theorem 5 of \cite{szekely2009brownian} 
and the Glivenko-Cantelli theorem. 
If $\Y$ is a univariate series, then 
the test statistic will also be asymptotically distribution-free.
 %

Applying this test, we find 
${\cal Q}_6(\widehat{\Y}, 12) = 30.92.$
By applying a resampling scheme similar to that in Section 2.6, we find this has a 
$p$-value $\approx 0.$ 
This indicates significant serial dependence in the series. 
To remove this dependence we fit a vector autoregression (VAR) of order three using ordinary least squares. 
Let $\widehat{\utwi{E}}$ denote the estimated residuals. 
We find 
${\cal Q}_6(\widehat{\utwi{E}}, 12) = 0.10$, with $p$-value $\approx 0.09$.
Hence, this simple VAR model is sufficient for removing all serial dependence in the series $\widehat{\Y},$
and no nonlinear modeling is necessary.

Given the test results above, we proceed under the assumption that the $\widehat{\utwi{E}}$ are iid,
and now apply our ICA methodology.
First we test whether the components of $\widehat{\utwi{E}}$ are ICs.
The ICs test statistic is 
${\cal U}_n(\widehat{\utwi{E}}) = 5.27,$ with $p$-value $\approx 0,$
 indicating significant dependence. 
To simplify parameter interpretation, the elements of $\widehat{\utwi{E}}$ are scaled by their standard deviations 
$(0.45, 0.59, 0.48, 0.41, 0.52, 0.65)'$.
Next, PCA was applied to obtain approximately uncorrelated components $\widehat{\utwi{Z}}.$
The ICs test statistic for these standardized PC scores is  
${\cal U}_n(\widehat{\utwi{Z}}) = 0.41,$ with $p$-value $\approx 0$, 
hence the PCs are not ICs. 
Finally, ICs $\widehat{\utwi{S}}$ are estimated  using the PITdCovICA method.
The ICs test statistic 
is 
${\cal U}_n(\widehat{\utwi{S}}) = -0.41,$ with $p$-value $\approx 0.85$,
hence we conclude that ICs do exist for the residuals $\widehat{\utwi{E}}$. 
These results are summarized in Table \ref{my.dcov}.
Note that linear transformation from $\widehat{\utwi{E}}$ to $\widehat{\utwi{Z}}$ did not induce any serial dependence and  
the transformation from $\widehat{\utwi{Z}}$ to $\widehat{\utwi{S}}$ was an orthogonal rotation, which distance covariance is invariant to, see Table \ref{my.auto.dcov}.
%

The estimated mixing matrix is shown in Table \ref{fig42}(a).
Since the components of $\widehat{\utwi{E}}$ have roughly the same variance, 
and since $\widehat{Cov}_n(\widehat{\utwi{S}}) = \bm{I}$, we have $\widehat{Cov}_n(\widehat{\utwi{E}}) \approx \widehat{\m}\widehat{\m}'.$ 
From this, we see that the sum of squares of the $k$th row of $\widehat{\m}$ gives the variance of $\widehat{\utwi{E}}_{k}$. Thus, the square of each element gives the {\it proportion} of the variance of $\widehat{\utwi{E}}_{k}$ explained by the ICs. In this view, we can remove the smaller coefficients to simplify the interpretation and find
%
CA: $\hat{e}_{1} = -0.79\hat{s}_{2} + 0.32\hat{s}_{3} -0.43\hat{s}_{4} -0.26\hat{s}_{5}$,
FL: $\hat{e}_{2} = -0.77 \hat{s}_{1} + 0.55\hat{s}_{3} + 0.22\hat{s}_{5}$, 
IL: $\hat{e}_{3} =  -0.41 \hat{s}_{1}   -0.32\hat{s}_{4}  + 0.84\hat{s}_{6}$, 
MI: $\hat{e}_{4} =  0.26\hat{s}_{2} +0.31\hat{s}_{3}  -0.91\hat{s}_{4}$,
OH: $\hat{e}_{5} =  -0.42 \hat{s}_{1} -0.61\hat{s}_{3} -0.60 \hat{s}_{4} -0.28 \hat{s}_{6}$, 
WI: $\hat{e}_{6} =  -0.29 \hat{s}_{1} -0.18 \hat{s}_{4} -0.92 \hat{s}_{5} $.

From these results, we see that $\hat{s}_{4}$ is related to each state. 
Time plots of $\hat{s}_{4}$ and the change series of seasonally
adjusted GDP shows a positive association. This supports
the hypothesis that $-\hat{s}_{4}$ is a national component of the unemployment rate.
The component $\hat{s}_{5}$ is largely specific to WI, while 
CA and OH have the most complicated structure. 
%

The estimated uncorrelating matrix used to estimate $\widehat{\utwi{Z}}$ is shown in Table \ref{fig42}(b).
%
The first component is roughly an equally weighted average of all six series.
The second component gives positive loadings to CA and FL, and negative loadings to the midwestern states. 
The inverse of the estimated mixing matrix is shown in Table \ref{fig42}(c).
Besides the third column, the remaining components give much more relative weight to individual states then the PC scores do.
We conclude that these six series can adequately be modeled by a vector autoregression, of which the 
errors can be decomposed into mutually 
independent components.

\section{Concluding Remarks}

In this paper, we extended the distance covariance dependence measure to develop a novel approach for ICA.
We estimated ICs using a nonparametric probability integral transformation with a generalized 
nonparametric whitening method that simultaneously minimizes all forms of dependence among the components. 
We established the limiting properties of the proposed estimator under weak regularity conditions and 
proposed a flexible resampling-based framework for statistical inference.  
In contrast with the existing literature, we proposed a test statistic and procedure for checking the existence of 
mutually independent components. The test procedure is consistent and is found to work well in simulation and real 
examples. Simulation results showed that the proposed approach to ICA outperforms the competing methods.
We then applied the proposed method to two real examples and obtained 
sensible interpretations for the data. These examples also highlighted the difference between ICA and PCA. 

There are several ways to extend the proposed ICA methods. We primarily considered 
the case of iid observations. However, many applications, especially in finance, have 
serially uncorrelated, but dependent data. Extension of the proposed approach to handle such data can substantially 
increase its applicability. Second, we only considered the lower dimensional applications in this paper. Many 
applications encounter high dimensional data. Developing an efficient estimation procedure for the proposed ICA methods
 to handle high dimensional data is challenging, but important. 
 Finally, adaptive methods for the proposed estimators may be considered for application of ICA to data which are only locally stationary. 
%

\section*{Appendix} 
\begin{appendix}
\renewcommand{\theequation}{A.\arabic{equation}}
\renewcommand{\thesubsection}{A.\arabic{subsection}}
\setcounter{equation}{0}
%
%


\baselineskip=24pt

\subsection*{Proof of Theorem \ref{theta.as}}


\subsubsection*{Lemma A.1}\label{thm1} 
\begin{it}
Under Assumptions 2.1 and 2.3,
$  \widetilde{{\cal J}}_n({\theta})  \stackrel{\mbox{a.s.}}{\longrightarrow} \widetilde{{\cal J}}({\theta})  $
as  $n {\rightarrow} \infty$, for any ${\theta} \in \Theta$.
\end{it}
\begin{proof}
\begin{eqnarray*}
|\widetilde{{\cal J}}_n({\theta}) - \widetilde{{\cal J}}({\theta})| 
 &=&\Big| \sum_{i=1}^{d-1} \,  \V_n(\widetilde{\utwi{U}}_k(\T), \widetilde{\utwi{U}}_{k^+}(\T)) - \V(U_k(\T), U_{k^+}(\T)) \Big| \allowdisplaybreaks[3] \\
 &\le& \sum_{i=1}^{d-1}  \, \Big| \V_n(\widetilde{\utwi{U}}_k(\T), \widetilde{\utwi{U}}_{k^+}(\T)) - \V(U_k(\T), U_{k^+}(\T)) \Big| \allowdisplaybreaks[3] \\
 &\le& \sum_{i=1}^{d-1}  \, \Big( \Big| \V_n(\widetilde{\utwi{U}}_k(\T), \widetilde{\utwi{U}}_{k^+}(\T)) -  \V_n({\utwi{U}}_k(\T),{\utwi{U}}_{k^+}(\T)) \Big|  \\ 
  & & \quad \quad +
  \Big| \V_n({\utwi{U}}_k(\T),{\utwi{U}}_{k^+}(\T)) -  \V(U_k(\T), U_{k^+}(\T)) \Big| \Big)
\end{eqnarray*}
for any ${\theta} \in \Theta$.
For each $k$, 
$\left| \V_n(\widetilde{\utwi{U}}_k(\T), \widetilde{\utwi{U}}_{k^+}(\T)) -  \V_n({\utwi{U}}_k(\T),{\utwi{U}}_{k^+}(\T)) \right| \stackrel{\mbox{a.s.}}{\longrightarrow} 0$ by Assumption 2.3 and the continuous mapping theorem, and 
$\left| \V_n({\utwi{U}}_k(\T),{\utwi{U}}_{k^+}(\T)) -  \V(U_k(\T), \utwi{U}_{k^+}(\T)) \right| \stackrel{\mbox{a.s.}}{\longrightarrow} 0$ by Assumption 2.1, 
the triangle inequality, H{\"o}lder's inequality,
the strong law of large numbers for $U$-statistics 
(see \citealp{hoeffding1961strong}), 
and Slutsky's theorem,
as $n {\rightarrow} \infty$, thus establishing the assertion. 
\end{proof}


Let ${\cal D}$ denote any metric on ${\cal SO}(d)$, continuous in its first argument, such that for all $\w, \A \in {\cal SO}(d)$, ${\cal D}(\w, \A) = 0$ if and only if there exists a $\p_{\pm}$ such that $\w = \p_{\pm}\A$, and ${\cal D}(\w, \A) > 0$ otherwise. Partition ${\cal SO}(d)$ into equivalence classes via ${\cal D}$: the ${\cal D}$-distance between any two elements within an equivalence class is 0, and the ${\cal D}$-distance between any two elements from different equivalence classes is greater than 0. Let ${\cal SO}(d)_{{\cal D}}$ be the quotient space ${\cal SO}(d)/{\cal D}$ of these equivalence classes. Then $\w = \A$ on ${\cal SO}(d)_{{\cal D}}$ if and only if ${\cal D}(\w, \A) = 0$.

\subsubsection*{Lemma A.2}\label{lemma3} 
\begin{it}
Under Assumptions 2.1 and 2.3, 
$\widetilde{\cal J}_n({\theta})$ is Lipschitz continuous for $\T : \w_{\T} \in {\cal SO}(d)_{{\cal D}}$.
%
%
\end{it}

\begin{proof}
First, note that the composition of two Lipschitz continuous functions is also Lipschitz continuous. 
$\utwi{S}_{\T}$ is a trigonometric compositions of Lipschitz functions with respect to ${\theta}$, hence it is Lipschitz continuous. 
Lipschitz continuity of $\widetilde{\utwi{U}}(\T)$ follows from Assumption 2.3.

To establish the Lipschitz continuity of $\widetilde{\cal J}_n({\theta})$ it is sufficient to show
$\V_n(\widetilde{\utwi{U}}_k(\T), \widetilde{\utwi{U}}_{k^+}(\T))$ is Lipschitz continuous for $k = 1,\ldots,d-1$.
%
%
%
The Euclidean norm is a Lipschitz function, as is a linear combinations of two Lipschitz functions.
The product of two bounded Lipschitz functions is a Lipschitz functions as well. 
It is clear that $\V_n(\widetilde{\utwi{U}}_k(\T), \widetilde{\utwi{U}}_{k^+}(\T))$ is uniformly bounded for a fixed dimension $d$. This establishes the Lipschitz continuity of $\widetilde{\cal J}_n({\theta})$.
\end{proof}


\subsubsection*{Lemma A.3}\label{thm2B} 
\begin{it}
Under Assumptions 2.1 and 2.3, 
\begin{eqnarray*}\label{thm3eqn}
\sup_{\T : W_{\T} \in {\cal SO}(d)_{{\cal D}}} |\widetilde{\cal J}_n(\T) - \widetilde{\cal J}(\T)|
   \stackrel{\mbox{a.s.}}{\longrightarrow} 0 
\quad \mbox{as} \quad n {\rightarrow} \infty. 
\end{eqnarray*}
\end{it}
%
\begin{proof}
Applying 
the Arzel\'{a}-Ascoli theorem from complex analysis it is sufficient to show: \\
(i) $\widetilde{\cal J}_n(\T) \stackrel{\mbox{a.s.}}{\longrightarrow} \widetilde{\cal J}(\T)$ for each $\T : \w_{\T} \in \Xi_0$, some countable dense subset of ${\cal SO}(d)_{{\cal D}}$, and \\
(ii) $\lim_{c \rightarrow \infty} \overline{\lim}_{n} \; m_{\frac{1}{c}}(\widetilde{\cal J}_n) \stackrel{\mbox{a.s.}}{=} 0$, in which 
\begin{eqnarray*}
m_{\frac{1}{c}}(\widetilde{\cal J}_n) = \sup\left\{|\widetilde{\cal J}_n(\T) - \widetilde{\cal J}_n(\Ti)|:
\w_{\Ti},\w_{\T} \in {\cal SO}(d)_{{\cal D}}, \left|\left| \w_{\Ti}-\w_{\T} \right|\right|_F < 1/c \right\}. 
 \end{eqnarray*}

${\cal SO}(d)_{{\cal D}}$ is separable since it is compact. Consequently, there exists a countable dense subset, say $\Xi_0$. Lemma A.1 implies that $\widetilde{\cal J}_n(\T) \stackrel{\mbox{a.s.}}{\longrightarrow} \widetilde{\cal J}(\T)$ as $n \rightarrow \infty,$ for each $\w_{\T} \in {\cal SO}(d)_{{\cal D}}$, and in particular for each $\w_{\T} \in \Xi_0$. 

Let $\tilde{\utwi{u}} = \widetilde{\utwi{U}}(\T)$, $\tilde{\utwi{v}} = \widetilde{\utwi{U}}(\Ti)$, ${\utwi{u}} = {\utwi{U}}(\T)$ and ${\utwi{v}} = {\utwi{U}}(\Ti)$. 
Lemma A.2 implies that there exists a constant $0<L<\infty$ such that for any $\w_{\Ti}, \w_{\T} \in {\cal SO}(d)_{{\cal D}}$, $\big| \big| \w_{\Ti} - \w_{\T} \big| \big|_F \le \delta_1$ implies 
$|\tilde{\utwi{u}}_i - \tilde{\utwi{v}}_i| < L \delta_1 $ for all $i = 1,\ldots,n$.
Note that
\begin{eqnarray*}
|\widetilde{{\cal J}}_n({\theta}) - \widetilde{\cal J}_n(\Ti) | 
 &=&\Big| \sum_{{\ell}=1}^{d-1} \,  \V_n(\widetilde{\utwi{U}}_{\ell}(\T), \widetilde{\utwi{U}}_{{\ell}^+}(\T)) - \V_n(\widetilde{\utwi{U}}_{\ell}(\Ti), \widetilde{\utwi{U}}_{{\ell}^+}(\Ti))  \Big| \allowdisplaybreaks[3] \\
 &\le& \sum_{{\ell}=1}^{d-1}  \, \Big| \V_n(\widetilde{\utwi{U}}_{\ell}(\T), \widetilde{\utwi{U}}_{{\ell}^+}(\T)) - \V_n(\widetilde{\utwi{U}}_{\ell}(\Ti), \widetilde{\utwi{U}}_{{\ell}^+}(\Ti)) \Big| \\
 &=& \sum_{{\ell}=1}^{d-1}  \, \Big| \left(T_{1,n}^{({\ell})}(\T) + T_{2,n}^{({\ell})}(\T) - T_{3,n}^{({\ell})}(\T) \right) - \left(T_{1,n}^{({\ell})}(\Ti) + T_{2,n}^{({\ell})}(\Ti) - T_{3,n}^{({\ell})}(\Ti) \right) \Big| \\
 &\le& \sum_{{\ell}=1}^{d-1}  \, \Big| T_{1,n}^{({\ell})}(\T) - T_{1,n}^{({\ell})}(\Ti) \Big| +\Big| T_{2,n}^{({\ell})}(\T) -  T_{2,n}^{({\ell})}(\Ti) \Big| + \Big|T_{3,n}^{({\ell})}(\T)) - T_{3,n}^{({\ell})}(\Ti)) \Big| ,
\end{eqnarray*}
in which the $T_{j,n}^{({\ell})}(\T)$ are defined as $T_{j,n}(\widetilde{\utwi{U}}_{\ell}(\T), \widetilde{\utwi{U}}_{{\ell}^+}(\T)),$ analogous to Equation (\ref{IXY}).
Applying standard Euclidean norm inequalities we note the following inequalities
\begin{footnotesize}
\begin{eqnarray*} 
\Big| T_{1,n}^{({\ell})}(\T) - T_{1,n}^{({\ell})}(\Ti) \Big|  &=&
\Bigg| {n \choose 2}^{-1} \sum_{i < j} | \tilde{u}_{i,\ell} - \tilde{u}_{j,\ell}  | |\tilde{\utwi{u}}_{i,\ell^+} - \tilde{\utwi{u}}_{j,\ell^+} |
- {n \choose 2}^{-1} \sum_{i < j} | \tilde{v}_{i,\ell} - \tilde{v}_{j,\ell}  | |\tilde{\utwi{v}}_{i,\ell^+} - \tilde{\utwi{v}}_{j,\ell^+} | \Bigg| \allowdisplaybreaks[3] \\
&\le&
{n \choose 2}^{-1} \sum_{i < j} \Big|| \tilde{u}_{i,\ell} - \tilde{u}_{j,\ell}  | |\tilde{\utwi{u}}_{i,\ell^+} - \tilde{\utwi{u}}_{j,\ell^+} |
- | \tilde{v}_{i,\ell} - \tilde{v}_{j,\ell}  | |\tilde{\utwi{v}}_{i,\ell^+} - \tilde{\utwi{v}}_{j,\ell^+} | \Big|  \allowdisplaybreaks[3] \\
&\le&
{n \choose 2}^{-1} \sum_{i < j} | \tilde{u}_{i,\ell} - \tilde{u}_{j,\ell}  | |(\tilde{\utwi{u}}_{i,\ell^+} - \tilde{\utwi{u}}_{j,\ell^+}) - (\tilde{\utwi{v}}_{i,\ell^+} - \tilde{\utwi{v}}_{j,\ell^+}) |  + \\
& &  {n \choose 2}^{-1} \sum_{i < j}
 | (\tilde{u}_{i,\ell} - \tilde{u}_{j,\ell}) -(\tilde{v}_{i,\ell} - \tilde{v}_{j,\ell})  | |\tilde{\utwi{v}}_{i,\ell^+} - \tilde{\utwi{v}}_{j,\ell^+} | \allowdisplaybreaks[3] \\
 &\le&
\Bigg( {n \choose 2}^{-1} \sum_{i < j} | \tilde{u}_{i,\ell} - \tilde{u}_{j,\ell}  | \Bigg) \Bigg( {n \choose 2}^{-1} \sum_{i < j} |(\tilde{\utwi{u}}_{i,\ell^+} - \tilde{\utwi{u}}_{j,\ell^+}) - (\tilde{\utwi{v}}_{i,\ell^+} - \tilde{\utwi{v}}_{j,\ell^+}) | \Bigg) + \\
& & \Bigg( {n \choose 2}^{-1} \sum_{i < j} 
 | (\tilde{u}_{i,\ell} - \tilde{u}_{j,\ell}) -(\tilde{v}_{i,\ell} - \tilde{v}_{j,\ell})  | \Bigg) \Bigg({n \choose 2}^{-1} \sum_{i < j} |\tilde{\utwi{v}}_{i,\ell^+} - \tilde{\utwi{v}}_{j,\ell^+} |  \Bigg) \allowdisplaybreaks[3] \\
  &\le&
\Bigg( {n \choose 2}^{-1} \sum_{i < j} | \tilde{u}_{i,\ell} - \tilde{u}_{j,\ell}  | \Bigg) \Bigg( {n \choose 2}^{-1} \sum_{i < j} ( |\tilde{\utwi{u}}_{i,\ell^+} - \tilde{\utwi{v}}_{i,\ell^+}| + |\tilde{\utwi{u}}_{j,\ell^+} - \tilde{\utwi{v}}_{j,\ell^+} | ) \Bigg) + \\
& &  \Bigg({n \choose 2}^{-1} \sum_{i < j} 
 (| \tilde{u}_{i,\ell} - \tilde{v}_{i,\ell} | + |\tilde{u}_{j,\ell} - \tilde{v}_{j,\ell} | ) \Bigg) \Bigg( {n \choose 2}^{-1} \sum_{i < j} |\tilde{\utwi{v}}_{i,\ell^+} - \tilde{\utwi{v}}_{j,\ell^+} | \Bigg)  \allowdisplaybreaks[3] \\
  &\le&
\Bigg( {n \choose 2}^{-1} \sum_{i < j} | \tilde{\utwi{u}}_{i}  - \tilde{\utwi{u}}_{j}   | \Bigg) \Bigg( {n \choose 2}^{-1} \sum_{i < j} ( |\tilde{\utwi{u}}_{i} - \tilde{\utwi{v}}_{i}| + |\tilde{\utwi{u}}_{j} - \tilde{\utwi{v}}_{j} | ) \Bigg) + \\
& &  \Bigg({n \choose 2}^{-1} \sum_{i < j} 
 ( |(\tilde{\utwi{u}}_{i} - \tilde{\utwi{v}}_{i}| + |\tilde{\utwi{u}}_{j} - \tilde{\utwi{v}}_{j}) | ) \Bigg) \Bigg( {n \choose 2}^{-1} \sum_{i < j}  |\tilde{\utwi{v}}_{i} - \tilde{\utwi{v}}_{j} | \Bigg), 
\end{eqnarray*}
%
%
\begin{eqnarray*}
\Big| T_{2,n}^{({\ell})}(\T) - T_{2,n}^{({\ell})}(\Ti) \Big|  &=&
\Bigg| {n \choose 2}^{-1} \sum_{i < j} | \tilde{u}_{i,\ell} - \tilde{u}_{j,\ell}  | {n \choose 2}^{-1} \sum_{i < j} |\tilde{\utwi{u}}_{i,\ell^+} - \tilde{\utwi{u}}_{j,\ell^+} | \\
& & - {n \choose 2}^{-1} \sum_{i < j} | \tilde{v}_{i,\ell} - \tilde{v}_{j,\ell}  | {n \choose 2}^{-1} \sum_{i < j} |\tilde{\utwi{v}}_{i,\ell^+} - \tilde{\utwi{v}}_{j,\ell^+} | \Bigg| \allowdisplaybreaks[3] \\
&\le& \Bigg| {n \choose 2}^{-1} \sum_{i < j} | \tilde{u}_{i,\ell} - \tilde{u}_{j,\ell}  | \Bigg| \Bigg| {n \choose 2}^{-1} \sum_{i < j} |\tilde{\utwi{u}}_{i,\ell^+} - \tilde{\utwi{u}}_{j,\ell^+} | - {n \choose 2}^{-1} \sum_{i < j} |\tilde{\utwi{v}}_{i,\ell^+} - \tilde{\utwi{v}}_{j,\ell^+} | \Bigg| \\
& & + \Bigg|{n \choose 2}^{-1} \sum_{i < j} | \tilde{u}_{i,\ell} - \tilde{u}_{j,\ell}  | - {n \choose 2}^{-1} \sum_{i < j} | \tilde{v}_{i,\ell} - \tilde{v}_{j,\ell}  |\Bigg| \Bigg| {n \choose 2}^{-1} \sum_{i < j} |\tilde{\utwi{v}}_{i,\ell^+} - \tilde{\utwi{v}}_{j,\ell^+} | \Bigg| \\
&\le& \Bigg( {n \choose 2}^{-1} \sum_{i < j} | \tilde{u}_{i,\ell} - \tilde{u}_{j,\ell}  | \Bigg) 
\Bigg( {n \choose 2}^{-1} \sum_{i < j} |(\tilde{\utwi{u}}_{i,\ell^+} - \tilde{\utwi{u}}_{j,\ell^+}) - (\tilde{\utwi{v}}_{i,\ell^+} - \tilde{\utwi{v}}_{j,\ell^+}) | \Bigg) \\
& & + \Bigg({n \choose 2}^{-1} \sum_{i < j} | (\tilde{u}_{i,\ell} - \tilde{u}_{j,\ell} ) - (\tilde{v}_{i,\ell} - \tilde{v}_{j,\ell} ) |\Bigg) \Bigg( {n \choose 2}^{-1} \sum_{i < j} |\tilde{\utwi{v}}_{i,\ell^+} - \tilde{\utwi{v}}_{j,\ell^+} | \Bigg) \allowdisplaybreaks[3] \\
&\le& \Bigg( {n \choose 2}^{-1} \sum_{i < j} | \tilde{u}_{i,\ell} - \tilde{u}_{j,\ell}  | \Bigg) 
\Bigg( {n \choose 2}^{-1} \sum_{i < j} (|\tilde{\utwi{u}}_{i,\ell^+} - \tilde{\utwi{v}}_{i,\ell^+}| + |\tilde{\utwi{u}}_{j,\ell^+} - \tilde{\utwi{v}}_{j,\ell^+} |) \Bigg) \\
& & + \Bigg({n \choose 2}^{-1} \sum_{i < j} (| \tilde{u}_{i,\ell} - \tilde{v}_{i,\ell} | + |\tilde{u}_{j,\ell} - \tilde{v}_{j,\ell} | ) \Bigg) \Bigg( {n \choose 2}^{-1} \sum_{i < j} |\tilde{\utwi{v}}_{i,\ell^+} - \tilde{\utwi{v}}_{j,\ell^+} | \Bigg) \allowdisplaybreaks[3] \\
&\le& \Bigg( {n \choose 2}^{-1} \sum_{i < j} | \tilde{\utwi{u}}_{i}  - \tilde{\utwi{u}}_{j}  | \Bigg) 
\Bigg( {n \choose 2}^{-1} \sum_{i < j}  ( |\tilde{\utwi{u}}_{i} - \tilde{\utwi{v}}_{i}| + |\tilde{\utwi{u}}_{j} - \tilde{\utwi{v}}_{j} | )  \Bigg) \\
& & + \Bigg({n \choose 2}^{-1} \sum_{i < j}  ( |\tilde{\utwi{u}}_{i} - \tilde{\utwi{v}}_{i}| + |\tilde{\utwi{u}}_{j} - \tilde{\utwi{v}}_{j} | )  \Bigg) \Bigg( {n \choose 2}^{-1} \sum_{i < j} |\tilde{\utwi{v}}_{i} - \tilde{\utwi{v}}_{j} | \Bigg), 
\end{eqnarray*}
%
\begin{eqnarray*}
& & \hspace{-2.5cm} \Bigg| {n \choose 3}^{-1} \sum_{i < j < k} | \tilde{u}_{i,\ell} - \tilde{u}_{j,\ell}  | |\tilde{\utwi{u}}_{i,\ell^+} - \tilde{\utwi{u}}_{k,\ell^+} |
- {n \choose 3}^{-1} \sum_{i < j < k} | \tilde{v}_{i,\ell} - \tilde{v}_{j,\ell}  | |\tilde{\utwi{v}}_{i,\ell^+} - \tilde{\utwi{v}}_{k,\ell^+} | \Bigg| \allowdisplaybreaks[3] \\
&\le&
{n \choose 3}^{-1} \sum_{i < j < k} \Big|| \tilde{u}_{i,\ell} - \tilde{u}_{j,\ell}  | |\tilde{\utwi{u}}_{i,\ell^+} - \tilde{\utwi{u}}_{k,\ell^+} |
- | \tilde{v}_{i,\ell} - \tilde{v}_{j,\ell}  | |\tilde{\utwi{v}}_{i,\ell^+} - \tilde{\utwi{v}}_{k,\ell^+} | \Big| \allowdisplaybreaks[3] \\
&\le&
{n \choose 3}^{-1} \sum_{i < j < k} | \tilde{u}_{i,\ell} - \tilde{u}_{j,\ell}  | |(\tilde{\utwi{u}}_{i,\ell^+} - \tilde{\utwi{u}}_{k,\ell^+}) - (\tilde{\utwi{v}}_{i,\ell^+} - \tilde{\utwi{v}}_{k,\ell^+}) |  + \\
& &  {n \choose 3}^{-1} \sum_{i < j < k}
 | (\tilde{u}_{i,\ell} - \tilde{u}_{j,\ell}) -(\tilde{v}_{i,\ell} - \tilde{v}_{j,\ell})  | |\tilde{\utwi{v}}_{i,\ell^+} - \tilde{\utwi{v}}_{k,\ell^+} |  \allowdisplaybreaks[3] \\
 &\le&
\Bigg( {n \choose 3}^{-1} \sum_{i < j < k} | \tilde{u}_{i,\ell} - \tilde{u}_{j,\ell}  | \Bigg) \Bigg( {n \choose 3}^{-1} \sum_{i < j < k} |(\tilde{\utwi{u}}_{i,\ell^+} - \tilde{\utwi{u}}_{k,\ell^+}) - (\tilde{\utwi{v}}_{i,\ell^+} - \tilde{\utwi{v}}_{k,\ell^+}) | \Bigg)  + \\
& &  \Bigg( {n \choose 3}^{-1} \sum_{i < j < k} 
 | (\tilde{u}_{i,\ell} - \tilde{u}_{j,\ell}) -(\tilde{v}_{i,\ell} - \tilde{v}_{j,\ell})  | \Bigg) \Bigg( {n \choose 3}^{-1} \sum_{i < j < k} |\tilde{\utwi{v}}_{i,\ell^+} - \tilde{\utwi{v}}_{k,\ell^+} |  \Bigg) \allowdisplaybreaks[3] \\
  &\le&
\Bigg( {n \choose 3}^{-1} \sum_{i < j < k} | \tilde{u}_{i,\ell} - \tilde{u}_{j,\ell}  | \Bigg) \Bigg( {n \choose 3}^{-1} \sum_{i < j < k} ( |\tilde{\utwi{u}}_{i,\ell^+} - \tilde{\utwi{v}}_{i,\ell^+}| + |\tilde{\utwi{u}}_{k,\ell^+} - \tilde{\utwi{v}}_{k,\ell^+} | ) \Bigg)+ \\
& &  \Bigg( {n \choose 3}^{-1} \sum_{i < j < k} 
 (| \tilde{u}_{i,\ell} - \tilde{v}_{i,\ell} | + |\tilde{u}_{j,\ell} - \tilde{v}_{j,\ell} | ) \Bigg) \Bigg( {n \choose 3}^{-1} \sum_{i < j < k} |\tilde{\utwi{v}}_{i,\ell^+} - \tilde{\utwi{v}}_{k,\ell^+} |  \Bigg) \\
   &\le&
\Bigg( {n \choose 3}^{-1} \sum_{i < j < k} | \tilde{\utwi{u}}_{i}  - \tilde{\utwi{u}}_{j}   | \Bigg) \Bigg( {n \choose 3}^{-1} \sum_{i < j < k}( |\tilde{\utwi{u}}_{i} - \tilde{\utwi{v}}_{i}| + |\tilde{\utwi{u}}_{k} - \tilde{\utwi{v}}_{k} | ) \Bigg)+ \\
& &  \Bigg( {n \choose 3}^{-1} \sum_{i < j < k} 
 ( |\tilde{\utwi{u}}_{i} - \tilde{\utwi{v}}_{i}| + |\tilde{\utwi{u}}_{j} - \tilde{\utwi{v}}_{j} | ) \Bigg) \Bigg( {n \choose 3}^{-1} \sum_{i < j < k} |\tilde{\utwi{v}}_{i} - \tilde{\utwi{v}}_{k} |  \Bigg), \\
 &\le& \Bigg( {n \choose 2}^{-1} \sum_{i < j} | \tilde{\utwi{u}}_{i}  - \tilde{\utwi{u}}_{j}  | \Bigg) 
\Bigg( {n \choose 2}^{-1} \sum_{i < j}  ( |\tilde{\utwi{u}}_{i} - \tilde{\utwi{v}}_{i}| + |\tilde{\utwi{u}}_{j} - \tilde{\utwi{v}}_{j} | )  \Bigg) \\
& & + \Bigg({n \choose 2}^{-1} \sum_{i < j}  ( |\tilde{\utwi{u}}_{i} - \tilde{\utwi{v}}_{i}| + |\tilde{\utwi{u}}_{j} - \tilde{\utwi{v}}_{j} | )  \Bigg) \Bigg( {n \choose 2}^{-1} \sum_{i < j} |\tilde{\utwi{v}}_{i} - \tilde{\utwi{v}}_{j} | \Bigg), 
\end{eqnarray*}
\end{footnotesize}
and similarly for the remaining terms in $T_{3,n}.$ Hence, 
\begin{footnotesize}
%
\begin{eqnarray*}
\Big| T_{3,n}^{({\ell})}(\T) - T_{3,n}^{({\ell})}(\Ti) \Big|  &\le& 2\Bigg( {n \choose 2}^{-1} \sum_{i < j} | \tilde{\utwi{u}}_{i}  - \tilde{\utwi{u}}_{j}  | \Bigg) 
\Bigg( {n \choose 2}^{-1} \sum_{i < j}  ( |\tilde{\utwi{u}}_{i} - \tilde{\utwi{v}}_{i}| + |\tilde{\utwi{u}}_{j} - \tilde{\utwi{v}}_{j} | )  \Bigg) \\
& & + 2\Bigg({n \choose 2}^{-1} \sum_{i < j}  ( |\tilde{\utwi{u}}_{i} - \tilde{\utwi{v}}_{i}| + |\tilde{\utwi{u}}_{j} - \tilde{\utwi{v}}_{j} | )  \Bigg) \Bigg( {n \choose 2}^{-1} \sum_{i < j} |\tilde{\utwi{v}}_{i} - \tilde{\utwi{v}}_{j} | \Bigg).
\end{eqnarray*}
\end{footnotesize}
Therefore, $\big| \big| \w_{\Ti} - \w_{\T} \big| \big|_F \le \delta_1$ implies 
\begin{footnotesize}
\begin{eqnarray*}
|\widetilde{{\cal J}}_n({\theta}) - \widetilde{\cal J}_n(\Ti) | 
 &\le& 4\sum_{{\ell}=1}^{d-1}  \, \left(
 \Bigg( {n \choose 2}^{-1} \sum_{i < j} | \tilde{\utwi{u}}_{i}  - \tilde{\utwi{u}}_{j}  | \Bigg) 
\Bigg( {n \choose 2}^{-1} \sum_{i < j}  ( |\tilde{\utwi{u}}_{i} - \tilde{\utwi{v}}_{i}| + |\tilde{\utwi{u}}_{j} - \tilde{\utwi{v}}_{j} | )  \Bigg) +  \right.  \\
& & \quad \quad \quad \left.  
 \Bigg({n \choose 2}^{-1} \sum_{i < j}  ( |\tilde{\utwi{u}}_{i} - \tilde{\utwi{v}}_{i}| + |\tilde{\utwi{u}}_{j} - \tilde{\utwi{v}}_{j} | )  \Bigg) \Bigg( {n \choose 2}^{-1} \sum_{i < j} |\tilde{\utwi{v}}_{i} - \tilde{\utwi{v}}_{j} | \Bigg)  \right) \allowdisplaybreaks[3]  \\ 
  &\le& 4\sum_{{\ell}=1}^{d-1} 
  \left( \frac{2}{n} \sum_{i = 1}^n  |\tilde{\utwi{u}}_{i} - \tilde{\utwi{v}}_{i}| \right)
   \left( {n \choose 2}^{-1} \sum_{i < j} | \tilde{\utwi{u}}_{i}  - \tilde{\utwi{u}}_{j}  |+ {n \choose 2}^{-1} \sum_{i < j} |\tilde{\utwi{v}}_{i} - \tilde{\utwi{v}}_{j} |  \right) \\
     &\le& 
   \left( {n \choose 2}^{-1} \sum_{i < j} | \tilde{\utwi{u}}_{i}  - \tilde{\utwi{u}}_{j}  |+{n \choose 2}^{-1} \sum_{i < j} |\tilde{\utwi{v}}_{i} - \tilde{\utwi{v}}_{j} |  \right) 8(d-1) L \delta_1 \\
       &=& B_n 8(d-1)  L \delta_1
\end{eqnarray*}
\end{footnotesize}
For each $c \in \mathbb{N}$, let $\delta = \min\{1/c, \delta_1\}$. Now observe that for $n \in \mathbb{N}$
\begin{eqnarray*}
m_{\delta}(\widetilde{\cal J}_n) =
\sup_{|| W_{\Ti} - W_{\T} ||_F < \delta} |\widetilde{{\cal J}}_n({\theta}) - \widetilde{\cal J}_n(\Ti) |  
 \le  B_n 8(d-1)  L \delta.
\end{eqnarray*}
Let $B = E|\utwi{u} - \utwi{u}'| + E|\utwi{v} - \utwi{v}'| $ in which $\utwi{u}'$ and $\utwi{v}'$ are iid copies of 
$\utwi{u}$ and $\utwi{v}$, respectively. 
By Assumption 2.1 we have $B < \infty$, and 
by the SLLN for $U$-statistics $B_n \stackrel{\mbox{a.s.}}{\longrightarrow} B$, as $n \rightarrow \infty$. 
Therefore,
$\overline{\lim}_{n} \, m_{\delta}(\widetilde{\cal J}_n) 
 \le
\overline{\lim}_{n} \,
B_n 8(d-1)  L \delta
 \stackrel{\mbox{a.s.}}{=}
 B 8(d-1)  L \delta .$
As $c \rightarrow \infty$, $\delta = \min\{1/c, \delta_1\} = 1/c$. Therefore, the claim is established by noting 
%
%

\hspace{+2.4cm}$
\lim_{c \rightarrow \infty} \overline{\lim}_{n} \; m_{\frac{1}{c}}(\widetilde{\cal J}_n) 
 \le_{a.s.} 
\lim_{c \rightarrow \infty} 
 \left( B 8(d-1)  L \right)/c  = 0. 
 $
\end{proof}
%
%
%
\subsubsection*{Proof of Theorem 2.4}\label{thm3} 
Under Assumptions 2.1 and 2.3, note that for any $n \in \mathbb{N}$,
$\widetilde{\cal J}_n(\T_0) 
\ge
\widetilde{\cal J}_n(\widetilde{\T}_n) $
and 
$\widetilde{\cal J}(\widetilde{\T}_n)
\ge
\widetilde{\cal J}(\T_0). $
Hence,
\begin{eqnarray*}
\widetilde{\cal J}_n(\T_0)  - \widetilde{\cal J}(\T_0)
\ge
\widetilde{\cal J}_n(\widetilde{\T}_n) - \widetilde{\cal J}(\T_0) 
\ge
\widetilde{\cal J}_n(\widetilde{\T}_n) -  \widetilde{\cal J}(\widetilde{\T}_n),
\end{eqnarray*}
and
\begin{eqnarray*}
|\widetilde{\cal J}_n(\widetilde{\T}_n) - \widetilde{\cal J}(\T_0)|
&\le& \max \big( 
|\widetilde{\cal J}_n(\T_0) - \widetilde{\cal J}(\T_0)|,
|\widetilde{\cal J}_n(\widetilde{\T}_n) - \widetilde{\cal J}(\widetilde{\T}_n)|
\big) \\
&\le&  \sup_{\T : W_{\T} \in {\cal SO}(d)_{{\cal D}}} |\widetilde{\cal J}_n(\T) - \widetilde{\cal J}(\T)|.
     \end{eqnarray*}


Therefore, Lemma A.3 implies that $\widetilde{\cal J}_n(\widetilde{\T}_n) \stackrel{\mbox{a.s.}}{\longrightarrow} \widetilde{\cal J}(\T_0) \; \mbox{as} \; n {\rightarrow} \infty$ for $\T : \w_{\T_0} \in {\cal SO}(d)_{{\cal D}}$. Note that the $\argmin$ mapping is continuous on ${\cal SO}(d)_{{\cal D}}$. Since ${\cal SO}(d)_{{\cal D}}$ is compact, the $\argmin$ of $\widetilde{\cal J}_n$ and $\widetilde{\cal J}$ exists in ${\cal SO}(d)_{{\cal D}}$; therefore, ${\w}_{\T_n} \stackrel{\mbox{a.s.}}{\longrightarrow}  \w _{\T_0}$, as $n {\rightarrow} \infty$, for $\w_{\T_0} \in {\cal SO}(d)_{{\cal D}}$. 
If ${\theta}_0 \in \overline{\Theta}$, in which $\overline{\Theta}$ is a sufficiently large compact subset of the space ${\Theta}$, then Lemma A.3 and the continuous mapping theorem imply $\widetilde{{\theta}}_n  \stackrel{\mbox{a.s.}}{\longrightarrow} {\theta}_0$ as $n {\rightarrow} \infty$. 
\qed

\baselineskip=18pt

\end{appendix}

\bibliographystyle{ECA_jasa}
\baselineskip=12pt

\bibliography{Matteson_ICA}


\clearpage\pagebreak\newpage
\thispagestyle{empty}

\begin{table}[h]
\caption{\label{sim} \baselineskip=10pt  Mean error distance, Equation (\ref{error}), approximate standard error, and average computation time in seconds (s) for $N = 1000$ simulations in $\mathbb{R}^4, \mathbb{R}^6,$ and $\mathbb{R}^8$ with sample size $n=1000$, by randomly selecting four of the 18 distributions shown in Figure \ref{AllSim}.}
\begin{center}
{\footnotesize 
\vspace{-0.0cm}
\begin{tabular}{|c|c|cccccc|}
\hline
  &  &  \multicolumn{2}{c}{Joint Estimation} & \multicolumn{2}{c}{Sequential Estimation} &  &    \\
 & ICA Method & dCovICA & {PITdCovICA} & dCovICA & {PITdCovICA} & FastICA & ProDenICA   \\
\hline
& Mean Error & 0.0739  & 0.0639 & 0.0864 & 0.0981  & 0.1879 & 0.0630  \\
$\mathbb{R}^4$ & Standard Error & 0.0016 & 0.0009 & 0.0019 & 0.0022 & 0.0048 & 0.0008   \\
&  Mean Time (s) & 9.52 & 5.45 &  1.79 & 3.19 & 0.02 & 3.33    \\
\hline
& Mean Error & 0.0834  & 0.0774 & 0.1192 & 0.1312  & 0.2719 & 0.0809  \\
$\mathbb{R}^6$ & Standard Error & 0.0009 & 0.0007 & 0.0018 & 0.0021 & 0.0052 & 0.0008   \\
&  Mean Time (s) & 16.51 &18.30 &  7.32 & 10.19 & 0.04 & 5.30    \\
\hline
& Mean Error & 0.0960  & 0.0841 & 0.1517 & 0.1600  & 0.3286 & 0.0954  \\
$\mathbb{R}^8$ & Standard Error & 0.0004 & 0.0004 & 0.0020 & 0.0020 & 0.0049 & 0.0008   \\
&  Mean Time (s) & 24.67 &26.97 &  16.99 & 21.84 & 0.05 & 7.09    \\
\hline
\end{tabular}
}
\end{center}
%
%
%
\vspace{1cm}
 \caption{\label{fig41} 
 \baselineskip=10pt ICA of the Freedman crime data, using the PITdCovICA estimator.
 The standardized observations $\widehat{\Y}$ consist of: the logarithm of population (total 1968, in thousands), nonwhite (percent nonwhite population, 1960), density (population per square mile, 1968), crime (crime rate per 100,000, 1969). 
 The fitted mixing matrix and its inverse are shown below. They define the relationship between the observations and the estimated ICs $\widehat{\utwi{S}}.$  
}
\centering
 \vspace*{0.2cm}
 {\footnotesize 
\begin{tabular}{|rrrr|rrrr|}
\multicolumn{4}{c}{$\widehat{\m}_n' : \widehat{\Y} = \widehat{\utwi{S}}\widehat{\m}_n'$} & \multicolumn{4}{c}{$\widehat{\m}_n'^{-1} : \widehat{\utwi{S}} = \widehat{\Y}\widehat{\m}_n'^{-1}$} \\
 \hline
 0.23  &  0.54 &   0.51   & 0.76 & -0.42  & -0.98  &  0.52  & -0.35 \\
 -0.72  & -0.31 &  -0.47  &  0.08  & 0.35  & -0.41  &  0.20  &  0.87 \\
  0.41  &  0.34  & -0.67  &  0.51  & 0.55 &  -0.23 &  -0.87  & -0.11\\
 -0.52  &  0.70  & -0.28  & -0.38  & 0.82  &  0.74 &   0.28 &  -0.44\\
\hline
 \end{tabular}}
 %
\vspace{1cm}
\caption{\label{my.dcov} \baselineskip=10pt 
Test statistic ${\cal U}_n(\cdot)$ (see Equation (\ref{test})), and approximate $p$-value (based on 1999 permutations) for joint test of mutually independent components for the seasonally adjusted monthly unemployment rates from January 1976 through August 2010 for $6$ states: 
$\widehat{\utwi{Y}}$ standardized observations; $\widehat{\utwi{E}}$ VAR(3) residuals; $\widehat{\utwi{Z}}$ estimated PCs from $\widehat{\utwi{E}}$; and $\widehat{\utwi{S}}$ estimated ICs from $\widehat{\utwi{E}}$.
}
\begin{center}
{\footnotesize   
\vspace{-0.2cm}
\begin{tabular}{|c|rrrr|}
\hline
${\cal U}_n(\cdot)$ & $\widehat{\utwi{Y}}$ & $\widehat{\utwi{E}}$ & $\widehat{\utwi{Z}}$ & $\widehat{\utwi{S}}$   \\
\hline
Test Statistic & 39.7 & 5.27 & 0.41 & -0.41 \\
Approx. $p$-value & 0 & 0 & 0 & 0.85  \\
\hline
\end{tabular}
}
\end{center}
\end{table}

\clearpage\pagebreak\newpage
\thispagestyle{empty}

\begin{table}[h]
\caption{\label{my.auto.dcov} \baselineskip=10pt 
Test statistic ${\cal Q}_d(\cdot, m)$, see Equation (\ref{Qtest}), and approximate $p$-value (based on 1999 permutations) for $m = 12$ lag joint test of multivariate serial dependence of the seasonally adjusted monthly unemployment rates from January 1976 through August 2010 for $d = 6$ states: 
$\widehat{\utwi{Y}}$ standardized observations; $\widehat{\utwi{E}}$ VAR(3) residuals; $\widehat{\utwi{Z}}$ estimated PCs from $\widehat{\utwi{E}}$; and $\widehat{\utwi{S}}$ estimated ICs from $\widehat{\utwi{E}}$.
}
\begin{center}
{\footnotesize 
\vspace{-0.2cm}
\begin{tabular}{|c|rrrr|}
\hline
${\cal Q}_6(\cdot, m = 12)$ & $\widehat{\utwi{Y}}$ & $\widehat{\utwi{E}}$ & $\widehat{\utwi{Z}}$ & $\widehat{\utwi{S}}$   \\
\hline
Test Statistic & 30.92 & 0.10  &   -0.02 & -0.02  \\
Approx. $p$-value & 0  & 0.09  & 0.54  & 0.54  \\
\hline
\end{tabular}
}
\end{center}
%
\vspace{1cm} 
%
 \caption{\label{fig42} 
 \baselineskip=12pt ICA of the standardized change in monthly unemployment rate percentage, using the PITdCovICA estimator.
 The standardized observations $\widehat{\utwi{E}}$ consist of state level unemployment for: CA, FL, IL, MI, OH, and WI. 
 These series were rescaled by $\widehat{\utwi{D}}$ to have unit variance. 
 The fitted mixing matrix and its inverse are shown below, along with the estimated uncorrelating matrix. 
 They define the relationship between the observations and the estimated PCs $\widehat{\utwi{Z}}$ and ICs $\widehat{\utwi{S}}.$  
}
\centering
 \vspace*{0.2cm}
 { \footnotesize 
\begin{tabular}{|rrrrrr|}
\multicolumn{6}{c}{(a) $\widehat{\m}_n' : \widehat{\utwi{E}} = \widehat{\utwi{S}}\widehat{\m}_n' \widehat{\utwi{D}}$} \\
 \hline
-0.15 & -0.77 & -0.41 & -0.03 & -0.42 & -0.29 \\
-0.79 & -0.09 & -0.08 & 0.26 & -0.11 &  0.16 \\
0.32 &  0.55 & -0.12 &  0.31 & -0.61 &  0.11 \\
-0.43 & 0.04 & -0.32 & -0.91 & -0.60 & -0.18 \\
-0.26 & 0.22 & 0.00 & 0.02 & -0.04 & -0.92 \\
-0.01 & -0.19 & 0.84 & 0.01 & -0.28 & -0.03 \\
\hline
 \multicolumn{6}{c}{(b) $\widehat{\utwi{O}}_n' : \widehat{\utwi{Z}} =\widehat{\utwi{E}} \widehat{\utwi{D}}^{-1} \widehat{\utwi{O}}_n' $} \\
 \hline
 -0.34 & 0.32 & -0.11 & 0.03 & 0.58 & 0.84 \\ 
  -0.19 & 0.76 & 0.17 & -0.23 & -0.15 & -0.62 \\ 
  -0.26 & -0.17 & 0.67 & 0.72 & 0.06 & -0.19 \\
 -0.34 & -0.19 & 0.09 & -0.34 & -0.91 & 0.38 \\
 -0.29 & -0.43 & 0.03 & -0.59 & 0.56 & -0.49 \\ 
 -0.27 & -0.07 & -0.79 & 0.50 & -0.15 & -0.39 \\
 \hline
  \multicolumn{6}{c}{(c) $\widehat{\m}_n'^{-1} : \widehat{\utwi{S}} = \widehat{\utwi{E}} \widehat{\utwi{D}}^{-1} \widehat{\m}_n'^{-1}$} \\
\hline
0.25 & -1.03 & 0.25 & -0.23 & -0.18 & -0.02 \\
-0.84 & 0.12 & 0.44 & 0.22 & 0.31 & -0.25 \\
-0.35 & -0.01 & -0.18 & -0.04 & 0.06 & 0.98 \\
0.31 & 0.50 &  0.48 & -0.79 & 0.21 & -0.03 \\
-0.45 & -0.11 & -0.82 & -0.30 & 0.10 & -0.46 \\
-0.24 & 0.34 & 0.08 & 0.11 & -0.96 & -0.03 \\
\hline
 \end{tabular}}
 \end{table}

\clearpage\pagebreak\newpage
\thispagestyle{empty}

\clearpage\pagebreak\newpage
\thispagestyle{empty}

%

\clearpage\pagebreak\newpage
\thispagestyle{empty}
%

\end{document}